\documentclass[apj]{emulateapj}

\newcommand{\be}{\begin{equation}}
\usepackage{threeparttable}
\usepackage{booktabs}
\newcommand{\ee}{\end{equation}}
\newcommand{\bea}{\begin{eqnarray}}
\newcommand{\eea}{\end{eqnarray}}

\usepackage{graphicx,times}
\usepackage{subfigure}
\usepackage{amsmath}
\usepackage{mathrsfs}
\usepackage{cases}
\usepackage{bm}
\usepackage{longtable}
\usepackage{hyperref}
\usepackage{epstopdf}
\usepackage{amsmath,bm}
\usepackage{amssymb}
\usepackage{natbib}
\usepackage{morefloats}
\usepackage{multirow}
\usepackage{array}
\usepackage{verbatim}
\usepackage[stable]{footmisc}

\begin{document}

\title{Turbulent dynamo in a conducting fluid and partially ionized gas}

\author{Siyao Xu\altaffilmark{1} and A. Lazarian\altaffilmark{2}}

\altaffiltext{1}{Department of Astronomy, School of Physics, Peking University, Beijing 100871, China; syxu@pku.edu.cn}
\altaffiltext{2}{Department of Astronomy, University of Wisconsin, 475 North Charter Street, Madison, WI 53706, USA; lazarian@astro.wisc.edu}

\begin{abstract}

By following the Kazantsev theory and taking into account both  
microscopic and turbulent diffusion of magnetic fields,
we develop a unified treatment of the kinematic and nonlinear stages of turbulent dynamo,
and study the dynamo process for a full range of magnetic Prandtl number $P_m$ and ionization fractions.
We find a striking similarity between the dependence of dynamo behavior on $P_m$ in a conducting fluid and 
$\mathcal{R}$ (a function of ionization fraction) in partially ionized gas. 
In a weakly ionized medium, 
the kinematic stage is largely extended, including not only exponential growth but a new regime of dynamo characterized by
{\it linear-in-time growth of magnetic field 
strength}, and the resulting magnetic energy is much higher than the kinetic energy carried by viscous-scale eddies. 
Unlike the kinematic stage, the subsequent nonlinear stage is unaffected by microscopic diffusion processes and 
has a universal {\it linear-in-time growth of magnetic energy} with the  
growth rate as a constant fraction $3/38$ of the turbulent energy transfer rate,
showing good agreement with earlier numerical results.
Applying the analysis to the first stars and galaxies, 
we find that the kinematic stage is able to generate a field strength only an order of magnitude smaller than 
the final saturation value.
But the generation of large-scale magnetic fields 
can only be accounted for by the relatively inefficient nonlinear stage
and requires longer time than the free-fall time. 
It suggests that magnetic fields may not have played a dynamically important role 
during the formation of the first stars. 
\end{abstract}


\section{Introduction}

Magnetic fields are observed to be ubiquitous in the universe
\citep{Rei12,Bec12,Ner13}, 
and play as a dynamically important element in many astrophysical systems.  
Growing evidence suggests that magnetic fields were already space-filling at early cosmic times and had strengths
in high-redshift galaxies comparable to local galaxies
\citep{Bern08, Mur09,Ham12}. 
The first cosmic seed fields, which can be a relic from the very early universe 
\citep{TurW88}
or alternatively generated by additional astrophysical mechanisms
\citep{Bierm50, Laz92,Sch03,Med04, Xu08},
are many orders of magnitude lower than the present-day field strength. 
In view of this it is important to answer questions regarding how further strong amplification of magnetic fields arises, 
what is their role for primordial star formation, and 
when dynamically important magnetic fields appeared in the first galaxies.

Both magnetic fields and turbulence are essential ingredients of the present-day picture of the interstellar medium 
(see e.g., \citealt{Dra11, XLY14, Xuc16, XZp}).
The most efficient process of magnetic field generation is identified as turbulent motions 
\citep{Batc50,Kaza68,KulA92}.
Turbulence in the early Universe was created during the gravitational structure formation of the first stars and galaxies, and its
presence has been confirmed by cosmological simulations 
(e.g., \citealt{Abel02,Grei08}). 
The consequent turbulent dynamo leads to an efficient exponential growth of magnetic field via stretching field lines by random velocity shear
\citep{Kaza68,KulA92,Bran05}. 
 Here we focus on the turbulent dynamo process 
on scales below the outer scale of turbulent motions.

In the weak field limit where the kinematic approximation holds, 
the standard theory for turbulent dynamo is the Kazantsev theory 
\citep{Kaza68}.
The dynamo instability exists for both spatially smooth viscous-scale motions and rough inertial-range turbulent velocities, 
and the Kazantsev theory is applicable to all scales up to the external scale of turbulence
\citep{Ru81,Nov83,Sub97,Vin01,SchK02,Bold04,Hau04,Bran05}.
When the magnetic energy becomes comparable to the turbulent kinetic energy of the smallest turbulent eddies, the velocity shear 
driven by these eddies is largely suppressed due to the strong magnetic back reaction.
As we will discuss in this paper, 
the kinematic approximation breaks down on scales below the scale where the magnetic energy reaches equipartition with the kinetic energy,
but is still valid over larger scales where the magnetic energy is 
overwhelmed by the kinetic energy. 
Importantly, the arising nonlinearities modify the resulting efficiency of the turbulent dynamo, but do not affect the dynamo process  
which is still dictated by the Kazantsev theory. 
In what follows, the stages of the turbulent dynamo with negligible and important magnetic back reaction are referred to as 
``kinematic stage" and ``nonlinear stage", respectively, 
bearing in mind that the kinematic approximation and Kazantsev theory stand on all scales in the former stage, but only on limited scales 
where the kinetic energy dominates over the magnetic energy in the latter stage.

It was the dynamo action in the kinematic stage that attracted ample attention,
whereas no satisfactory analytical description of the nonlinear stage has been developed
\citep{Sc02}.
Until recently, understanding on the nonlinear stage has been dramatically advanced by direct numerical experiments, 
which is found to be characterized by a linear growth of magnetic energy in time in the case of Kolmogorov turbulence until the equipartition with the turbulence forcing
(see \citealt{CVB09, BJL09, Bere11} and \citealt{BL15} for a review).
According to the numerical results provided in 
\citet{CVB09, BJL09, Bere11}, 
magnetohydrodynamic (MHD) turbulence emerges on scales smaller than 
the equipartition scale in the nonlinear stage.
The efficiency of the growth of magnetic energy, 
which is defined as the ratio between magnetic energy growth rate and hydrodynamic energy transfer rate, 
was found to be a universal constant with a numerically measured value much smaller than unity.
This finding is in striking contrast to earlier theoretical considerations in e.g. 
\citet{Sch02},
where the efficiency is assumed to be of order unity. 
The updated numerical results suggest that at the scale corresponding to the 
equipartition between magnetic and kinetic energies, the stretching effect is mostly cancelled out by turbulent diffusion. 
We caution that the term ``turbulent diffusion'' used in this paper refers to the diffusion of 
magnetic fields mediated by turbulent magnetic reconnection
\citep{LV99}, 
which is intrinsically related to the process of Richardson diffusion. 
\footnote{
Richardson diffusion 
\citep{Rich26}
was initially introduced for hydrodynamic turbulence and is fully consistent with the Kolmogorov theory of turbulence.
The explosive separation of magnetic field lines in MHD turbulence conforms to Richardson diffusion, 
which implies the breakdown of the flux-conservation constraint in MHD turbulence and can be used to recover the 
\citet{LV99}
theory on turbulent reconnection 
\citep{Ey10,Eyink2011}.}
The turbulent diffusion of dynamically important magnetic fields was
termed as ``reconnection diffusion" in
\citet{La05}.
Its significance on the star-formation process is supported by numerical simulations 
\citep{Sam10, LiM15, Gon16}
and observations  
\citep{LaE12}.
The ``turbulent diffusion" used in this paper is interchangeable with the ``reconnection diffusion".

Besides the turbulent diffusion of magnetic fields,
there exist other diffusion processes associated with microscopic plasma physics, such as resistive diffusion and ion-neutral collisional damping (ambipolar diffusion),
which are also related to the dissipation, or equivalently, damping, of the magnetic-fluctuation energy. 
The presence of the microscopic diffusion of magnetic fields 
can modify the turbulent dynamo model that was 
established and numerically tested in the conditions with high magnetic Prandtl number $P_m$ and high ionization degree 
\citep{KulA92, Sub97}.
Meanwhile,
numerical studies on low-$P_m$ and low-ionization turbulent dynamo are challenging and entail high computational costs. 
Most existing simulations are restricted to a limited range of $P_m$ near unity, 
while nature features either large or small $P_m$ 
\citep{SchK02,Rob00}.
Therefore, it is necessary for the analysis of the turbulent dynamo to cover a wide range of physical parameters. 
In this work we will refer to the existing simulations, while the whole range of our theoretical predictions is expected to be tested by future numerical efforts.

Our goal is to investigate the turbulent dynamo process during both the kinematic and nonlinear stages, and
further achieve the generalization of the turbulent dynamo over a full range of $P_m$ and ionization fractions. 
We follow the Kazantsev theory to describe the distribution of passive magnetic fluctuations in the kinematic regime, 
and take into account both turbulent diffusion and microscopic diffusion of magnetic fields.
As an important application of our analytical results, we examine the turbulent dynamo action during the formation of the first stars and galaxies,
and estimate the timescales for the kinematic and nonlinear stages to generate large-scale and strong magnetic fields.  
We also carry out a comparison between our results and those obtained in earlier works, e.g., 
\citet{SchoSch12}, \citet{Schob13}.

The paper is organized as follows. 
Physical insight and formulation of the kinematic and nonlinear stages of turbulent dynamo are presented in Section 2. 
Detailed analysis of different evolutionary stages of magnetic energy is performed 
in Section 3 for a conducting fluid and Section 4 for partially ionized gas. 
In particular, we describe the properties of the MHD turbulence related to the dynamo action in the nonlinear stage in Section 5. 
Next, in Section 6, the analytical results in the case of partially ionized gas are applied to the formation of the first stars and galaxies. 
Comparison with earlier works and further discussions are provided in Section 7. 
Finally, we summarize the main results in Section 8.

\section{Kinematic and nonlinear stages of turbulent dynamo}

\subsection{The Kazantsev theory of turbulent dynamo}
\label{ssec: kanzt}

When the initially weak magnetic field is introduced in a turbulent flow, 
magnetic field lines that are assumed to be frozen into turbulent plasma flow are lengthened due to the random stretching/shearing
driven by turbulent eddies 
\citep{Batc50}.
The rate of the line-stretching action is determined by the turnover rate of turbulent eddies, 
which can be derived by following the Kolmogorov theory for describing 
the two-point statistics of hydrodynamic incompressible turbulence. 
In the inertial range of turbulence cascade spanning from the energy injection scale $L$ to the viscous scale $1/k_\nu$,
the turbulent velocity at wavenumber $k$ is 
\begin{equation}\label{eq: scallaw}
  v_k = V_L (Lk)^{-\frac{1}{3}}, 
\end{equation}
and the eddy turnover rate at $k$ is 
\begin{equation}
    \Gamma = v_k k =  L^{-\frac{1}{3}} V_L k^{\frac{2}{3}},
\end{equation}
where $V_L$ is the turbulent velocity at $L$.

In the kinematic regime when the magnetic energy is relatively small compared to the turbulent kinetic energy,
the theory of linear turbulent dynamo was introduced by 
\citet{Kaza68}.
In the framework of Kazantsev dynamo in Fourier space, magnetic energy extends over a spectrum 
\begin{equation}
       \mathcal{E}=\frac{1}{2}  V_A^2     =\frac{1}{2} \int_0^{k^\prime} M(k,t) dk, 
\end{equation}
where $V_A=B/\sqrt{4\pi\rho}$ is Alfv\'{e}n speed. 
The Kazantsev spectrum of magnetic energy has a dependence on both time $t$ and $k$, scaling as $\sim k^{3/2}$ 
\citep{Kaza68, KulA92, Sch02, Bran05, Feder11, XuH11},
\begin{equation}\label{eq: mspgrw}
   M(k,t) = M_0 \exp{\bigg(\frac{3}{4}  \int \Gamma dt \bigg)} \bigg( \frac{k}{k_\nu} \bigg)^\frac{3}{2},
\end{equation}
where it is assumed that the initial magnetic energy $\mathcal{E}_0$ is concentrated on 
the scale of the smallest eddy size $k_\nu$, 
and $M_0 = \mathcal{E}_0 / k_\nu$. 
At the current scale $k^\prime$,
the magnetic energy of interest is from a larger-scale magnetic field at $k<k^\prime$, 
which appears as a uniform background field with respect to the $k^\prime$ mode of magnetic fluctuations 
\citep{KulA92}, 
whereas the smaller-scale magnetic fields are dynamically irrelevant.

The dynamo instability exists in both smooth velocity field at the viscous scale and rough flows over the turbulent scales 
\citep{Kaza68,Ru81,Nov83,Sub97,Vin01,SchK02,Bold04}.
\citet{Sub97}
extended the dynamo study for the case with single-scale velocity 
(e.g., \citealt{Ze83, Kle86, Ruz89})
to the context where dynamo is driven by turbulent velocities over a range of characteristic scales, 
and showed that the critical magnetic Reynolds number for excitation of a mode extending to an arbitrary scale within the inertial range of turbulence 
is the same as in the case when the velocity has a single spatial scale. 
The numerical study by 
\citet{Hau04}
showed that the magnetic spectrum over the turbulence inertial range is in qualitative agreement with the Kazantsev slope  
when the magnetic field is weak. 
They numerically confirmed the applicability of the Kazantsev theory in the kinematic regime 
for both high $P_m$ and $P_m =1$ cases.

In view of these theoretical arguments and numerical evidence, we perform our calculations based on the Kazantsev theory of turbulent 
dynamo in both the sub-viscous and inertial ranges of turbulent velocities in the kinematic regime. 
The actual dynamo growth of magnetic energy is determined by the competition between the stretching and microscopic diffusion of magnetic fields 
in the kinematic stage, or the turbulent diffusion of magnetic fields in the nonlinear stage.
In both the kinematic and nonlinear stages, the turbulent eddies on the scales falling in the kinematic regime act in isolation and 
amplify the magnetic fields in the same manner complying with the Kazantsev theory.

\subsection{Kinematic stage of turbulent dynamo}

In the kinematic stage, the magnetic energy is smaller than the turbulent kinetic energy of the viscous-scale eddies, and the 
turbulent diffusion of magnetic fields are not involved in the dynamo process.
The viscous-scale eddies have the fastest eddy turnover rate and dominantly drive the 
kinematic dynamo at a rate, 
\begin{equation}\label{eq: grnu}
  \Gamma = \Gamma_\nu = v_{k_\nu} k_\nu=L^{-\frac{1}{3}}V_Lk_\nu^\frac{2}{3}. 
\end{equation}
Here $v_{k_\nu}$ is the turbulent velocity at $k_\nu$.
In an ideal situation when the magnetic energy dissipation (i.e.microscopic diffusion of magnetic fields) is absent, 
the magnetic energy evolves exponentially as 
\begin{equation}\label{eq: kinexp}
     \frac{d \mathcal{E}}{d t}= 2 \Gamma_\nu \mathcal{E}.
\end{equation}
It is well known that in this case the growth of magnetic energy is accompanied by a fast transfer of the bulk of magnetic energy toward smaller scales
\citep{Sch02}. 
The magnetic energy spectrum given in Eq. \eqref{eq: mspgrw} takes a simpler form 
\begin{equation}\label{eq: mesm}
   M(k,t) = M_0 \exp{\bigg(\frac{3}{4} \Gamma_\nu t \bigg)} \bigg( \frac{k}{k_\nu} \bigg)^\frac{3}{2}. 
\end{equation}
The peak scale of $M(k,t)$ varies as 
\begin{equation}\label{eq: kpdif}
     k_p = k_\nu \exp{ \Big( \frac{3}{5} \Gamma_\nu t \Big)}. 
\end{equation}
Studying numerically the evolution of magnetic energy, 
\citet{Schek02,Sc02} 
introduced external forcing on the viscous scale.
This is equivalent to the situation that the advecting fluid flows are driven by 
the smallest turbulent eddies which are critically damped by viscosity.
Within these simulations the inertial range of turbulent velocity is absent,
but it is still possible to investigate the geometrical structure of the fluctuating magnetic fields produced by the kinematic dynamo 
over the broad sub-viscous range in high-$P_m$ medium. They found 
the magnetic field lines possess a folding structure, with rapid transverse direction reversals, 
but basically no change of the parallel scale of magnetic-field variation up to the scale of the flow 
(also see \citealt{Ot98,Kin00}).

However, 
in the presence of significant magnetic energy dissipation, the above paradigm of the kinematic stage of dynamo is substantially  
modified,
which we will discuss in detail in the following sections. 
Since the time evolution of magnetic energy spectrum in the kinematic stage strongly depends on magnetic dissipation/microscopic diffusion, 
we will separately study the dynamo process in a conducting fluid and partially ionized gas (Sections \ref{sec: fulion} and \ref{sec: partiond}), 
where the magnetic energy dissipation is dominated by different mechanisms.
In addition, 
as the magnetic energy grows and the significance of magnetic dissipation changes, 
we further subdivide the kinematic stage into multiple evolutionary stages with different efficiencies of magnetic field growth.

\subsection{Nonlinear stage of turbulent dynamo}\label{ssec: nondyn}

When the equipartition between the magnetic energy and the turbulent energy of the smallest eddies is achieved,
the magnetic back-reaction is strong enough to suppress the shear motions of these eddies. 
Consequently, the next larger-scale eddies which carry higher turbulent energy take over the dynamo action until the new equipartition sets in. 
In view of the theoretical studies by e.g.
\citet{Bie51, Kuls97, Sub99, Sch02}
and numerical simulations by 
\citet{Bran05, CVB09, BJL09, Bere11}, 
we consider that
accompanied with the process of achieving scale-by-scale equipartition, the dynamo growth  
proceeds to the nonlinear stage until reaching the full equipartition with the largest energy-containing eddies.

This stage of turbulent dynamo is nonlinear in terms of the strong back-reaction of magnetic fields on the turbulent eddies over scales 
below the equipartition scale. 
At every step in the nonlinear stage, over scales larger than the energy equipartition scale, 
the kinetic energy dominates over the magnetic energy and the turbulent motions are hydrodynamic.
The turbulent eddies remain in the kinematic regime and act to amplify the magnetic fields in accordance with the Kazantsev theory.
Whereas over smaller scales where the turbulent kinetic energy is in balance with the magnetic energy, 
the nonlinearities become important and kinematic approximation breaks down.  
On such scales, the turbulent eddies are unable to further amplify the magnetic field and are irrelevant in dynamo,
and the turbulent motions are modified to become Alfv\'{e}nic turbulence with the properties described by the 
\citet{GS95}
model of MHD turbulence.
\footnote{This feature can also be understood from a different perspective called frequency mismatching 
\citep{KulA92,Kuls97}.
For the magnetic fluctuations at scales smaller than the equipartition scale, 
their Alfv\'{e}n frequencies $kV_A$ exceed and mismatch with the turnover rate of the equipartition-scale eddies.
As a result, growth of magnetic energy at these scales is no longer possible. \\
The numerical testing of the 
\citet{GS95} 
model of MHD turbulence was influenced by the simulations that suffer from the bottleneck effect
\citep{BL10}.
The recent high-resolution MHD simulations in 
\citet{Bere14}
confirmed the 
\citet{GS95}
scaling.}
Notice that in this work 
we refer to ``MHD turbulence" as Alfv\'{e}nic turbulence with coherent magnetic fields over the characteristic scales of turbulent velocities.
The turbulent diffusion of magnetic fields arises in the MHD turbulence which is present over scales up to the equipartition scale.

The kinetic energy drained from the hydrodynamic cascade at the equipartition scale partially converts to the growing magnetic energy, 
which spreads over larger scales above the equipartition scale following the Kazantsev spectrum.
The remaining kinetic energy is passed down to smaller scales through the energy cascade of MHD turbulence, 
which exhibits a magnetic spectrum following the Kolmogorov $-5/3$ law
\citep{GS95}.
The turbulent diffusion of magnetic fields in MHD turbulence limits the efficiency of the dynamo process. 
The resulting magnetic energy spectrum contains both the ascending Kazantsev spectrum and descending MHD spectrum, 
and peaks at the evolving equipartition scale. 
Direct numerical evidence of this spectral form can be found in 
\citet{Bran05},
where the magnetic energy spectrum from their simulations 
can be fitted by the $k^{3/2}$ Kazantsev law at larger scales, and shows the $k^{-5/3}$ scaling at smaller scales (see Section \ref{sec: fulion}).
Accordingly, by applying the well-established Kazantsev theory in the kinematic regime on scales larger than the equipartition scale, 
and in the meantime taking into account the turbulent diffusion of magnetic fields on smaller scales, 
we next analytically derive the evolution law of magnetic energy during the nonlinear stage.

Unlike in the kinematic stage where only the microscopic magnetic diffusion exists and magnetic fields can be treated as frozen in 
the turbulent plasma on scales larger than the magnetic energy dissipation scale, 
in the nonlinear stage, MHD turbulence is developed and fast turbulent reconnection operates
\citep{LV99}.
As a result, in the MHD turbulence regime the frozen-in condition is not fulfilled and magnetic fields exhibit spontaneous stochasticity 
(see \citealt{Ey10, Eyink2011, Ey13} and \citealt{La15} for a review).
The violation of the frozen-in condition entails
the turbulent diffusion of magnetic fields and allows for a self-consistent treatment of the nonlinear stage.
During the nonlinear stage, both microscopic and turbulent diffusion of magnetic fields exist. 
But the microscopic diffusion operates at a rate much 
smaller than the turbulent diffusion rate over the length scales larger than the magnetic energy dissipation scale,
and thus can be neglected.

\subsubsection{Derivation of the dynamo efficiency in the nonlinear stage of turbulent dynamo}\label{sssec: dernon}

Magnetic energy concentrates at the spectral peak $k_p$ 
and the magnetic energy at larger scales $\mathcal{E} = 1/2  \int_0 ^ {k_p} M(k,t)  dk$ 
is in equipartition with the turbulent energy at $k_p$,
\begin{equation}\label{eq: enenli}
\mathcal{E}=\frac{1}{2} v_p^2 = \frac{1}{2} L^{-\frac{2}{3}} V_L^2 k_p^{-\frac{2}{3}} . 
\end{equation}
The dominant contribution for the shear motions comes from the turbulent eddies at the peak scale, i.e., equipartition scale. 
These are the smallest hydrodynamic eddies, which have the fastest turnover rate in comparison with larger-scale hydrodynamic eddies.
Their eddy turnover rate is 
\begin{equation}\label{eq: gamnol}
   \Gamma = \Gamma_p= L^{-\frac{1}{3}}V_Lk_p^\frac{2}{3}.
\end{equation}
The Kolmogorov scaling for hydrodynamic turbulence given by Eq. \eqref{eq: scallaw} 
is used in Eq. \eqref{eq: enenli} and \eqref{eq: gamnol}.
From the above equations we can easily find that the product of $\Gamma$ and $\mathcal{E}$ does not depend on $k_p$, namely,
\begin{equation}\label{eq: consg}
    \Gamma \mathcal{E}= \frac{1}{2} L^{-1} V_L^3 = \frac{1}{2} \epsilon, 
\end{equation}
where $\epsilon = v_k^3 k $ is a scale-independent constant within the Kolmogorov theory and 
conventionally defined in the literature as hydrodynamic energy transfer rate or Kolmogorov energy flux. 
It indicates that there is no energy dissipation along the turbulent cascade.

The growing magnetic energy is equal to the integral of the Kazantsev spectrum (Eq. \eqref{eq: mspgrw}) over $k<k_p$.
As discussed earlier, only the 
magnetic field on scales larger than the size of the eddies responsible for stretching, i.e., $k<k_p$, 
is relevant to the dynamo growth. 
It acts as a uniform background field with respect to these eddies. 
Meanwhile, the smaller-scale magnetic field is dynamically unimportant in the competition between the stretching and Lorentz tension.
By using Eq. \eqref{eq: mspgrw}, the growing magnetic energy is 
\begin{equation}\label{eq: nonllars}
\begin{aligned}
     \mathcal{E}   
                           &= \frac{1}{2} \int_0 ^ {k_p} M(k,t)  dk \\
                           &= \frac{1}{5}  \mathcal{E}_0 \exp{\bigg(\frac{3}{4} \int \Gamma dt \bigg)} \Big(\frac{k_p}{k_\nu}\Big)^\frac{5}{2}. 
\end{aligned}
\end{equation}

Next by applying $d \ln / dt$ to both sides of Eq. \eqref{eq: enenli}, we can compute 
\begin{equation} \label{eq: mantoene1}
    \frac{d\ln \mathcal{E}}{dt} = -\frac{2}{3} \frac{d \ln k_p}{dt}. 
\end{equation}
The same manipulation to Eq. \eqref{eq: nonllars} yields 
\begin{equation} \label{eq: mantoene2}
    \frac{d\ln \mathcal{E}}{dt} = \frac{3}{4} \Gamma + \frac{5}{2} \frac{d \ln k_p}{dt}. 
\end{equation}
Since the second term on the right-hand side in the above equation is negative, 
it is evident that the actual growth rate of magnetic energy is smaller than the hydrodynamic energy transfer rate. 
A combination of Eq. \eqref{eq: mantoene1} and \eqref{eq: mantoene2} leads to 
\begin{equation}\label{eq: intenon}
   \frac{d \ln \mathcal{E}}{ dt} = \frac{3}{19} \Gamma.
\end{equation}
By inserting the relation in Eq. \eqref{eq: consg}, we get 
\begin{equation}\label{eq: combjl}
   \frac{d \mathcal{E}}{ dt} = \frac{3}{38} \epsilon. 
\end{equation}
Since $\epsilon$ is a constant, it indicates that the magnetic energy in the nonlinear stage grows linearly with time. 
The ratio $3/38$ is determined by the scalings of both Kolmogorov and Kazantsev spectra, and 
reflects the fraction of turbulent energy that contributes to the actual growth of magnetic energy.
Thus, approximately, we can have 
\begin{equation}\label{eq: apnoen}
    \mathcal{E}  \sim \frac{3}{38} \epsilon t
\end{equation}
as the time evolution of magnetic energy, and based on the relation between $k_p$ and $\mathcal{E}$ from Eq. \eqref{eq: enenli}, we get 
from the above expression
\begin{equation}\label{eq: nonlkpap}
   k_p \sim \Big(\frac{1}{2}\Big)^\frac{3}{2} \Big(\frac{3}{38}\Big)^{-\frac{3}{2}} \epsilon^{-\frac{1}{2}} t^{-\frac{3}{2}}
\end{equation}
as the time evolution of the spectral peak $k_p$.

\subsubsection{Comparison with earlier works}

The evolution law for the magnetic energy in the nonlinear stage was earlier formulated by 
\citet{Sch02} 
as 
\begin{equation}\label{eq: Sch}
    \frac{d \mathcal{E}}{ dt}  \simeq \chi \epsilon - 2 \eta k^2_\text{rms} (t)  \mathcal{E}, 
\end{equation}
where $\chi$ is a constant of order unity, $\eta$ is the resistivity, and 
\begin{equation}
   k^2_\text{rms} (t)  = \frac{1}{\mathcal{E}} \int_0^\infty M(k,t) k^2 dk
\end{equation}
according to their definition. 

For comparison with the formula derived in this work, 
we combine Eq. \eqref{eq: mantoene1}, \eqref{eq: mantoene2}, and use the expressions given by
Eq. \eqref{eq: consg} and \eqref{eq: combjl}, to get 
\begin{equation}\label{eq: thisw}
      \frac{d  \mathcal{E}}{ dt} = \frac{3}{8} \epsilon - \frac{45}{152}  \epsilon.
\end{equation}
By comparing the two terms on the right hand side of Eq. \eqref{eq: Sch} given by 
\citet{Sch02} 
and Eq. \eqref{eq: thisw} from our result, we find the first difference is that in our treatment 
the constant $\chi$ is obviously less than unity. 
More importantly, our results show that for the nonlinear stage, the resistive term, i.e., the second term in Eq. \eqref{eq: Sch}, 
is negligibly small compared to the second term in Eq. \eqref{eq: thisw} which originates from 
the turbulent diffusion and is unrelated to the microscopic magnetic diffusivity.
The resistive diffusion only becomes comparably important as the turbulent diffusion at the small resistive scale.
As we discussed above, at each equipartition scale, only a small fraction of the kinetic energy is accumulated in the magnetic energy 
reservoir over larger scales,  
while the rest is transferred down to smaller scales via the cascade of MHD turbulence.
In the MHD turbulence over smaller scales, the stretching and diffusion of magnetic fields both occur at the eddy turnover rate
\citep{LV99, La05}.
Due to the cancellation between these two competing effects, there is no dynamo growth of magnetic energy over smaller scales, 
and thus the growing magnetic energy peaks at the equipartition scale, which also increases with time (Eq. \eqref{eq: nonlkpap}).

The turbulent diffusion of magnetic fields was also disregarded in 
\citet{KulA92}. 
Instead, they only 
considered the ambipolar diffusion of magnetic fields in the case of partially ionized gas, 
and used the ambipolar diffusion damping scale instead of the equipartition scale as the peak scale of the Kazantsev spectrum. 
Consequently, although the similar algebraic manipulations as shown in Section \ref{sssec: dernon} were carried out in 
\citet{KulA92}, 
they derived a higher efficiency of dynamo during the nonlinear stage. 
This theoretical expectation is disfavored by the numerical results presented in e.g. 
\citet{Bran05, CVB09, BJL09, Bere11}.

During the nonlinear stage, as the microscopic diffusion associated with plasma parameters is negligibly small, 
the turbulent diffusion dominates the magnetic field diffusion. As a result,
the growth of magnetic energy conforms to a universal evolution law dictated by turbulence properties, 
and the efficiency of dynamo is rather low, as shown in Eq. \eqref{eq: combjl}.
The linear-in-time growth of magnetic energy in the nonlinear stage was observed in numerical studies 
by, e.g. \citet{CVB09, BJL09, Bere11}. 
The scalings given by Eq. \eqref{eq: apnoen} and Eq. \eqref{eq: nonlkpap} correspond to the scalings established
in these numerical calculations.\footnote{The evolution of $k_p$ was discussed in e.g.\citet{BJL09} in terms of the change of equipartition scale in turbulent shock precursor dynamo. Our study above provides the analytical derivation from the first principles.} 
In particular, the evolution of magnetic energy in the nonlinear stage was expressed as
\begin{equation}\label{eq: adeff}
   \frac{d \mathcal{E}}{ dt} = A_d \epsilon
\end{equation}
in \citet{BJL09}, 
where $A_d$ represents the dynamo efficiency. 
The numerically measured values for $A_d$ vary 
from $0.04-0.05$ in
\citet{Bere11}
to $0.07$ in
\citet{CVB09}.
This universal efficiency of the conversion of turbulent energy to magnetic energy was also used by 
\citet{MB15}
as a fundamental parameter when studying the energy hierarchy in the intra-cluster medium, 
where the numerically evaluated value of $A_d$ from 
\citet{Bere11}
was adopted. 
In comparison with Eq. \eqref{eq: combjl}, our analytically derived value corresponding to $A_d$ is $3/38 \approx 0.08$. 
Taking into account the uncertainty
of the numerical results and an approximate nature of our scaling arguments, we consider the correspondence
with numerics as encouraging
(Jungyeon Cho, private communication).
The constant value of $A_d$ that we analytically obtained provides the physical justification of the 
earlier numerical finding 
that the nonlinear stage has a universal and much less-than-unity efficiency of amplifying magnetic field.

Hence
we present a unified treatment of both the kinematic and nonlinear stages of turbulent dynamo
as a competition between turbulent stretching and magnetic field diffusion. 
The advent of turbulent diffusion in the nonlinear stage dramatically decreases the efficiency 
of dynamo and changes the behavior of magnetic energy growth. 
Both the linear dependence on time and the small growth rate originate from the intrinsic properties of MHD turbulence.

\section{Turbulent dynamo in a conducting fluid }\label{sec: fulion}

In the limited case of fully ionized gas, we consider the resistive diffusion as the dominant energy dissipation effect. 
The ordinary Spitzer resistivity is 
\citep{Spit56}.
\begin{equation}
    \eta_s=\frac{c^2}{4\pi\sigma}  \sim \frac{1}{4} c^2 m_e^\frac{1}{2} Z e^2 \ln \Lambda (k_B T)^{-\frac{3}{2}}, 
\end{equation}
where $\sigma$ is the electric conductivity, and $\ln \Lambda$ is the Coulomb logarithm. 
On the other hand, the kinematic viscosity is determined by the Coulomb interaction between ions
\citep{Braginskii:1965},
\begin{equation}\label{eq: nui}
   \nu_i\sim\frac{c_{si}}{n_i \sigma_{ii}} \sim \frac{(k_BT)^\frac{5}{2}}{\pi n_i m_i^\frac{1}{2}  Z^4 e^4 \ln \Lambda},
\end{equation}
where $\sigma_{ii}$ is the cross-section of ion-ion Coulomb interaction, and $c_{si}$ is the sound speed in ions. 
The viscous cutoff $k_\nu$ of hydrodynamic turbulence
corresponds to the intersection between the hydrodynamic cascading rate $v_k k$ and viscous damping rate 
$k^2 \nu_i$. Using the Kolmogorov scaling (Eq. \eqref{eq: scallaw}), $v_k k = k^2 \nu_i$ gives the viscous scale,
\begin{equation}\label{eq: knui}
 k_\nu=L^{-\frac{1}{4}}V_L^{\frac{3}{4}}\nu_i^{-\frac{3}{4}}.
\end{equation}

Notice that as the magnetic field is strengthened, the viscosity becomes anisotropic. But since the viscosity parallel to magnetic field remains the same 
as the kinematic viscosity and overwhelms its perpendicular counterpart 
\citep{Schob13},
we adopt a constant viscosity as in Eq. \eqref{eq: nui} during the turbulent dynamo.
\footnote{As pointed out in 
\citet{GS06, sins07}, 
due to the reduction of the viscosity perpendicular to magnetic field, the magnetic field structure formed in the sub-viscous region 
may not be preserved.}
The relative importance between viscosity and resistivity can be referred to as the magnetic Prandtl number, which is defined as 
\begin{equation}
   P_m=\frac{\nu}{\eta}.
\end{equation}
We next discuss the evolution of magnetic energy at different ranges of $P_m$.

\subsection{Low magnetic Prandtl number ($P_m \leq 1$)}

At low $P_m$, the magnetic resistive scale lies inside the turbulent inertial range
($P_m<1$) or on the viscous cutoff($P_m=1$), 
namely, $k_R\leq k_\nu$. Since the magnetic energy beyond $k_R$ is 
dissipated resistively, the turbulent eddies on $k_R$ are responsible for the turbulent dynamo. 
The growth rate is characterized by the turnover rate of the resistive-scale eddies, 
\begin{equation}\label{eq: gammr}
   \Gamma\sim \Gamma_R=L^{-\frac{1}{3}}V_Lk_R^\frac{2}{3}.
\end{equation}
Meanwhile, the equalization between the growth rate and damping rate due to resistive dissipation $\Gamma = k_R^2 \eta_s$ is also satisfied at 
$k_R$. So the resistive scale is expected to be 
\begin{equation}\label{eq: lpkrdr}
   k_R= L^{-\frac{1}{4}} V_L^\frac{3}{4} \eta_s^{-\frac{3}{4}}. 
\end{equation}
Therefore, the ratio between $k_\nu$ and $k_R$ scales with $P_m$ as 
(also see \citealt{Moff61})
\begin{equation}\label{eq: knurpl}
   \frac{k_\nu}{k_R} = P_m^{-\frac{3}{4}},
\end{equation}
and thus $k_R= P_m^{3/4} k_\nu$.

Starting from the resistive scale, magnetic fluctuations can only spread out toward larger scales, but are suppressed over the sub-resistive region.  
It turns out the magnetic energy spectrum stays peaked at $k_R$ in the kinematic stage. 
So the magnetic energy grows as 
\begin{equation}\label{eq: drlpene}
\begin{aligned}
     \mathcal{E} &= \frac{1}{2} \int_0 ^ {k_R} M_0 \exp{\bigg(\frac{3}{4} \Gamma_R t \bigg)} \bigg( \frac{k}{k_R} \bigg)^\frac{3}{2} dk \\
                        &=\frac{1}{5} \mathcal{E}_0 \exp{\bigg(\frac{3}{4} \Gamma_R t \bigg)} , 
\end{aligned}
\end{equation}
where we define $M_0=\mathcal{E}_0 /k_R$. 
The kinematic saturation can be fulfilled at the balance between $\mathcal{E}$ in Eq. \eqref{eq: drlpene} and the turbulent energy at the resistive scale 
$E_{k,R}$. 
The corresponding saturated magnetic energy is 
\begin{equation}\label{eq: ekrlp}
    \mathcal{E}_\text{cr} = E_{k,R} = \frac{1}{2} L^{-\frac{2}{3}}V_L^2 k_R ^{-\frac{2}{3}} = P_m^{-\frac{1}{2}} E_{k,\nu},
\end{equation}
where 
\begin{equation}\label{eq: kinenu}
    E_{k,\nu}=\frac{1}{2} v_{k_\nu}^2 =\frac{1}{2} L^{-\frac{2}{3}}V_L^2 k_\nu ^{-\frac{2}{3}}
\end{equation}
is the turbulent energy at the viscous scale. 
By inserting $\mathcal{E}=\mathcal{E}_\text{cr}$ into Eq. \eqref{eq: drlpene}, we find the time interval for the kinematic stage, 
\begin{equation}\label{eq: crtlp}
    t_\text{cr} =  \frac{4}{3} \Gamma_R^{-1} \ln \Big( \frac{5E_{k,R}}{\mathcal{E}_0} \Big). 
\end{equation}
The above kinematic stage is subject to a severe damping effect due to significant resistivity, and thus referred to as damping stage, 
which is equivalent to the kinematic stage in the case of conducting fluid at $P_m<1$.

The nonlinear stage ensues following the damping stage. As discussed earlier, the magnetic energy during the nonlinear stage grows in a 
universal manner, independent of the dissipation mechanism. 
We apply the critical energy (Eq. \eqref{eq: ekrlp}) and critical time (Eq. \eqref{eq: crtlp}) as the 
boundary condition to Eq. \eqref{eq: combjl} and get the expression 
\begin{equation}\label{eq: ennoncr}
    \mathcal{E} = \mathcal{E}_\text{cr} + \frac{3}{38} \epsilon (t-t_\text{cr}). 
\end{equation}
By combining the above equation with Eq. \eqref{eq: enenli} and \eqref{eq: consg}, we further obtain 
\begin{equation}\label{eq: evokpcri}
    k_p = \Big[k_\text{cr}^{-\frac{2}{3}} + \frac{3}{19} \epsilon^\frac{1}{3}(t-t_\text{cr})\Big]^{-\frac{3}{2}},
\end{equation}
where the critical spectral peak $k_\text{cr}$ corresponding to $\mathcal{E}_\text{cr}$ is given by $k_R$ in this case.
The magnetic energy grows at a linear rate until the nonlinear saturation is achieved, where the magnetic energy 
is equal to the kinetic energy of the outer-scale turbulent eddy,
\begin{equation}\label{eq: satnlf}
    \mathcal{E}_\text{sat,nl} = \frac{1}{2} V_L^2. 
\end{equation}
The time required for $\mathcal{E}=\mathcal{E}_\text{sat,nl}$ is given by 
(Eq. \eqref{eq: gammr}, \eqref{eq: knurpl}, \eqref{eq: ekrlp}, and \eqref{eq: ennoncr})
\begin{equation}\label{eq: t2lp}
    t_2=\frac{19}{3} \Big(\frac{L}{V_L} - \Gamma_R^{-1} \Big) +t_\text{cr} = \frac{19}{3} \Big(\frac{L}{V_L} - P_m^{-\frac{1}{2}} \Gamma_\nu^{-1} \Big) +t_\text{cr}.
\end{equation}
Thus the duration of the nonlinear stage is 
\begin{equation}\label{eq: tsnllp}
     \tau_\text{nl}=\frac{19}{3} \Big(\frac{L}{V_L} - P_m^{-\frac{1}{2}} \Gamma_\nu^{-1} \Big),
\end{equation}
which is shortened under the condition of low $P_m$.

\subsection{High magnetic Prandtl number ($P_m > 1$)}

When $P_m$ is larger than unity, the viscous scale is larger than the resistive scale. 
It is necessary to point out that in our analysis, $P_m$ is not restricted to be near unity. 
In the case of $P_m>1$, the value of $P_m$ can range from $\gtrsim 1 $ to $\gg 1$.
Accordingly, the resistive scale can be comparably large or negligibly small compared to the viscous scale.
The dynamo action in the kinematic stage is driven by the viscous-scale eddies.
The growth rate $\Gamma_\nu$ is expressed as in Eq. \eqref{eq: grnu}, with the viscosity in ions involved (Eq. \eqref{eq: nui}).
Based on the simulations by 
\citet{CLV_newregime, CLV03}
and theoretical arguments by 
\citet{LVC04},
we assume that the magnetic fluctuations are not damped at the viscous cutoff of the hydrodynamic turbulent motions 
and can be developed in the viscosity-dominated range below the viscous scale.
We naturally assume that at the resistive scale, the dynamo growth rate is in equilibrium with the resistive dissipation rate, i.e., 
a statistically steady state between the line-stretching and resistive dissipation processes.
Equaling the damping rate due to resistive dissipation $k^2 \eta_s$ with the growth rate $\Gamma_\nu$ yields the expression of the resistive scale,
\begin{equation}\label{eq: hpkr}
    k_R= \sqrt{\frac{\Gamma_\nu}{\eta_s}} = L^{-\frac{1}{4}} V_L^\frac{3}{4} \nu_i^{-\frac{1}{4}} \eta_s^{-\frac{1}{2}},
\end{equation}
where the magnetic energy spectrum is cut off. 
From Eq. \eqref{eq: knui} and \eqref{eq: hpkr}, we find there exists 
(see also \citealt{Sch04})
\begin{equation}\label{eq: knrrat}
   \frac{k_\nu}{k_R} = \Big(\frac{\nu_i}{\eta_s}\Big)^{-\frac{1}{2}} = P_m^{-\frac{1}{2}}, 
\end{equation}
and hence $k_R=P_m^{1/2} k_\nu$.

The seed magnetic field is still assumed to reside at the smallest undamped eddy scale, i.e., $k_\nu$ in this case, with an initial energy $\mathcal{E}_0$. Starting from the seed field, we next analyze the dynamo growth of magnetic energy through the
evolutionary sequence.

(1) {\it Dissipation-free stage}

At the beginning of the kinematic stage, following the Kazantsev theory in the kinematic regime, 
the magnetic energy spectrum extends through the sub-viscous range but with the spectral peak far from the 
resistive scale. 
The magnetic energy grows exponentially (Eq. \eqref{eq: kinexp}),
\begin{equation} \label{eq: enedif}
      \mathcal{E}= \mathcal{E}_0 \exp{( 2 \Gamma_\nu t)}. 
\end{equation}
If the magnetic energy can grow up to $E_{k,\nu}$ before the energy spectrum peaks at $k_R$, 
the kinematic saturation occurs in the dissipation-free stage, and 
the equipartition between $\mathcal{E}$ in above equation and $E_{k,\nu}$ sets the corresponding time, 
\begin{equation}\label{eq: dfrtcr}
   t_\text{sat,k} =\frac{1}{2} \Gamma_\nu^{-1} \ln \Big(\frac{E_{k,\nu}}{\mathcal{E}_0}\Big).
\end{equation}
By substituting for the time from Eq. \eqref{eq: dfrtcr}, the spectral peak given in Eq. \eqref{eq: kpdif} reads
\begin{equation}\label{eq: hrdfkps}
   k_p (t_\text{sat,k}) = \Big(\frac{E_{k,\nu}}{\mathcal{E}_0}\Big)^\frac{3}{10} k_\nu.
\end{equation}
Otherwise, as the spectral peak propagates toward ever-smaller scales and reaches the resistive scale, 
the evolving Kazantsev spectrum of magnetic energy is cut off and remains peaked at the resistive scale, below which 
magnetic fluctuations are suppressed due to significant resistive diffusion. Therefore 
the kinematic stage proceeds to the viscous stage.

(2) {\it Viscous stage}

In the viscous stage, magnetic energy evolves according to 
\begin{equation}\label{eq: hpevis}
\begin{aligned}
     \mathcal{E} &= \frac{1}{2} \int_0 ^ {k_R} M_0 \exp{\bigg(\frac{3}{4} \Gamma_\nu t \bigg)} \bigg( \frac{k}{k_\nu} \bigg)^\frac{3}{2} dk \\
                           &=\frac{1}{5}   P_m^\frac{5}{4} \mathcal{E}_0 \exp{\bigg(\frac{3}{4} \Gamma_\nu t \bigg)} ,
\end{aligned}
\end{equation}
where the relation in Eq. \eqref{eq: knrrat} is used. 
This $3\Gamma_\nu/4$ growth rate was also pointed out in
\citet{KulA92}.

The transition time between the dissipation-free and viscous stages is set by equaling the above 
expression with Eq. \eqref{eq: enedif},
\begin{equation}\label{eq: hpt12}
   t_{12}=\Gamma_\nu^{-1} \ln \Big[\Big(\frac{1}{5}\Big)^\frac{4}{5} P_m\Big], 
\end{equation}
and the magnetic energy reached at $t_{12}$ is 
\begin{equation}\label{eq: hpet12}
   \mathcal{E}(t_{12})= \Big(\frac{1}{5}\Big)^\frac{8}{5} P_m^2 \mathcal{E}_0.
\end{equation}

We can see 
the condition for the viscous stage to be absent is $t_\text{sat,k}\leq t_{12}$, yielding (Eq. \eqref{eq: dfrtcr}, \eqref{eq: hpt12})
\begin{equation}
  P_m \geq 5^\frac{4}{5} \Big(\frac{E_{k,\nu}}{\mathcal{E}_0}\Big)^\frac{1}{2}.
\end{equation}
In contrast, at a smaller $P_m$, 
the equalization $\mathcal{E}=E_{k,\nu}$ in the viscous stage gives the saturation time of the kinematic stage
(Eq. \eqref{eq: hpevis})
\begin{equation}\label{eq: hpspcrt}
   t_\text{sat,k}= \frac{4}{3} \Gamma_\nu^{-1} \ln \Big[5 P_m^{-\frac{5}{4}} \Big(\frac{E_{k,\nu}}{\mathcal{E}_0}\Big)\Big].
\end{equation}
Therefore, the time interval of the viscous stage is 
\begin{equation}\label{eq: hptauvis}
   \tau_\text{vis} =  t_\text{sat,k}-t_{12}=\frac{4}{3} \Gamma_\nu^{-1} \ln \Big[5^{\frac{8}{5}}P_m^{-2}  \Big(\frac{E_{k,\nu}}{\mathcal{E}_0}\Big)\Big].
\end{equation}

(3) {\it Transitional stage}

At the kinematic saturation,  
magnetic energy is predominantly accumulated in the sub-viscous range, i.e., $k_p>k_\nu$. 
In fact, when the magnetic energy 
\begin{equation}
       \mathcal{E}=\frac{1}{2} \int_0^{k_p} M(k,t) dk
\end{equation}
becomes comparable to the kinetic energy of the viscous-scale eddies $E_{k,\nu}$, 
nonlinear effects intervene and suppress the growth of modes at $k>k_p$.  
Meanwhile the growth of modes at $k<k_p$ proceeds according to the same Kazantsev law.
As a result, the bulk of the magnetic energy propagates toward the viscous scale with the Kazantsev spectrum deformed and 
the advancing direction of the spectral peak reversed. 

We next calculate the spectral form left behind the evolving peak scale.
Over the larger scales away from resistive scale, 
the magnetic energy dissipation is insignificant.
The magnetic energy equal to the integral of the Kazantsev spectrum over $k < k_p$
(see Section \ref{ssec: kanzt})
is thus conserved and equalized with $E_{k,\nu}$ during the transitional stage, 
\begin{equation}\label{eq: contep}
\begin{aligned}
     \mathcal{E}   &= \frac{1}{2} \int_0 ^ {k_p} M_0 \exp{\bigg(\frac{3}{4} \Gamma_\nu t \bigg)} \bigg( \frac{k}{k_\nu} \bigg)^\frac{3}{2} dk \\
                          & =\frac{1}{5} \mathcal{E}_0 \exp{\bigg(\frac{3}{4} \Gamma_\nu t \bigg)}  \Big(\frac{k_p}{k_\nu} \Big)^\frac{5}{2} \\
                          & = E_{k,\nu}.
\end{aligned}
\end{equation}
Making use of the above relation, we get the time dependence of the spectral peak,
\begin{equation}\label{eq: fnonkp}
    k_p = \Big(\frac{5E_{k,\nu}}{\mathcal{E}_0}\Big)^\frac{2}{5} k_\nu \exp{\Big(-\frac{3}{10} \Gamma_\nu t \Big)}, 
\end{equation}
which gradually moves toward larger scales. The modes behind the spectral peak at $k>k_p$ no longer grow
because the Lorentz tension counterbalances the stretching action of the velocity shear,
while the modes at $k<k_p$ continue to grow in the same manner as mandated by
Eq. \eqref{eq: mesm}. 
Accordingly, the magnetic energy density at $k_p$ is
\begin{equation}
  M(k_p(t),t) = M_0 \exp{\bigg(\frac{3}{4} \Gamma_\nu t \bigg)} \bigg( \frac{k_p(t)}{k_\nu} \bigg)^\frac{3}{2}.
\end{equation}
Inserting the expression of $k_p$ from Eq. \eqref{eq: fnonkp} in the above equation gives 
\begin{equation}
   M(k_p(t),t) = \frac{\mathcal{E}_0}{k_\nu} \Big(\frac{5E_{k,\nu}}{\mathcal{E}_0}\Big)^\frac{3}{5} \exp{\Big(\frac{3}{10} \Gamma_\nu t \Big)},
\end{equation}
where $M_0$ is replaced by $\mathcal{E}_0 / k_\nu$. 
Combining with Eq. \eqref{eq: fnonkp}, the above equation can be reformulated in terms of $k_p$, 
\begin{equation}
    M(k_p(t)) = 5 E_{k,\nu} k_p(t)^{-1}. 
\end{equation}
During the transitional stage, the spectral peak moves from a sub-viscous scale to the viscous scale. 
As the peak scale increases, the spectrum behind it (higher-$k$) which corresponds to the peak scales 
at earlier time turns into a stationary state.
The magnetic field on scales smaller than the increasing peak scale should be arranged in a pattern satisfying the balance between the 
velocity shear and magnetic tension,
without either further bending or unwinding of field lines. 
This state can correspond to a folding structure of the magnetic field 
\citep{Cat97, Ot98, Cat99, Sc02, Schek02}.
It follows that the developed energy spectrum with the form
\begin{equation}\label{eq: negspec}
    M(k) = 5 E_{k,\nu} k^{-1}
\end{equation}
spreads out from the initial peak scale in the sub-viscous range at the beginning of the transitional stage 
up to the viscous scale at the end of the transitional stage. 
The negative spectral slope $-1$ is consistent with the 
conserved magnetic energy that we consider during the transitional stage.
This $k^{-1}$ tail below the viscous cutoff has been observed in  
numerical simulations on small-scale dynamo at $P_m>1$ 
\citep{Hau04}.

The above analysis shows that after $t_\text{sat,k}$, the magnetic energy at $k<k_p$ remains at the saturation level with 
$\mathcal{E}_\text{cr} = E_{k,\nu}$, 
while the spectral peak residing in the sub-viscous region moves up to the viscous scale following Eq. \eqref{eq: fnonkp}. 
Until $k_p$ reaches $k_\nu$, namely, 
the magnetic energy at $k<k_\nu$ is in equipartition with the kinetic energy of the viscous-scale eddies,
and the magnetic fields are spatially coherent at the viscous scale,
the nonlinear stage of turbulent dynamo is initiated. 
By equaling $k_p$ in Eq. \eqref{eq: fnonkp} with $k_\nu$, or equivalently, by 
\begin{equation}
    \mathcal{E}=\frac{1}{2} \int_0^{k_\nu} M(k,t) dk  = \frac{1}{5} \mathcal{E}_0 \exp{\bigg(\frac{3}{4} \Gamma_\nu t \bigg)}  = E_{k,\nu}, 
\end{equation}
we get the critical time for the onset of nonlinear stage, 
\begin{equation}\label{eq: tnlhesv}
    t_\text{cr} = \frac{4}{3} \Gamma_\nu^{-1} \ln \Big(\frac{5E_{k,\nu}}{\mathcal{E}_0}\Big). 
\end{equation}
From the time for the kinematic saturation $t_\text{sat,k}$ to $t_\text{cr}$, the transitional stage 
undergoes a period of time (Eq. \eqref{eq: dfrtcr}, \eqref{eq: tnlhesv})
\begin{equation}\label{eq: trantlr}
   \tau_\text{tran} = t_\text{cr} - t_\text{sat,k} = \Gamma_\nu^{-1} \ln \Big[ 5^\frac{4}{3} \Big(\frac{E_{k,\nu}}{\mathcal{E}_0}\Big)^\frac{5}{6} \Big]
\end{equation}
at $P_m \geq 5^{4/5} \sqrt{E_{k,\nu}/\mathcal{E}_0}$, and (Eq. \eqref{eq: hpspcrt}, \eqref{eq: tnlhesv})
\begin{equation}\label{eq: ttranlp}
   \tau_\text{tran} =  \frac{5}{3} \Gamma_\nu^{-1} \ln P_m
\end{equation}
at lower $P_m$.

The transitional stage emerges because of the 
nonlinear modification of magnetic energy spectrum at small scales. 
At this stage, the magnetic field is dynamically important in the sub-viscous range. 
Once the spectrum peaks at the viscous scale, the nonlinearity is activated inside the inertial range 
and generates MHD turbulence. 
Then the turbulent diffusion of magnetic fields comes into play.

(4) {\it Nonlinear stage~~~~}

Unlike the kinematic stage which includes the dissipation-free, (viscous), and transitional stages in the case of $P_m>1$ and 
is sensitive to the microscopic resistive diffusion,
during the subsequent nonlinear stage, the growth of magnetic energy is dictated by the universal expression 
Eq. \eqref{eq: ennoncr} until the nonlinear saturation. 
With the same critical energy $\mathcal{E}_\text{cr}$ (Eq. \eqref{eq: kinenu}) and critical time $t_\text{cr}$ (Eq. \eqref{eq: tnlhesv}), both scenarios with and without the viscous stage at $P_m >1$
have the same timescale for the nonlinear stage (Eq. \eqref{eq: ennoncr}),
\begin{equation}\label{eq: nontdrl}
    \tau_\text{nl}=\frac{19}{3} \Big(\frac{L}{V_L}-\Gamma_\nu^{-1}\Big), 
\end{equation}
and for the entire turbulent dynamo process,
\begin{equation}\label{eq: rltfsa}
     t_\text{sat,nl} = \frac{19}{3} \Big(\frac{L}{V_L}-\Gamma_\nu^{-1}\Big) + \frac{4}{3} \Gamma_\nu^{-1} \ln \Big(\frac{5E_{k,\nu}}{\mathcal{E}_0}\Big).
\end{equation}
After $t_\text{sat,nl}$, the final equipartition between the magnetic energy and the turbulent energy of the largest turbulent eddy
(Eq. \eqref{eq: satnlf}) is reached.

\subsection{Comparison between the turbulent dynamos at $P_m\leq1$ and $P_m>1$}

We next compare the growth timescales in the low and high $P_m$ cases. 
The evolutionary stages prior to the nonlinear stage all belong to the kinematic stage. 
Due to larger saturated energy ($E_{k,R}>E_{k,\nu})$ and lower growth rate ($\Gamma_R< \Gamma_\nu$),
we find that it takes longer time for the kinematic dynamo to achieve saturation at $k_R$ than at $k_\nu$, by a time difference 
(Eq. \eqref{eq: crtlp}, \eqref{eq: tnlhesv})
\begin{equation}
    \Delta t_\text{kin} = (P_m^{-\frac{1}{2}}-1) t_\text{cr} +\frac{4}{3} P_m^{-\frac{1}{2}} \Gamma_\nu^{-1} \ln (P_m^{-\frac{1}{2}}),
\end{equation}
where $P_m<1$ and $t_{cr}$ is the critical time at $P_m>1$ (Eq. \eqref{eq: tnlhesv}). 
Evidently, the nonlinear stage in the low-$P_m$ case is relatively short. The difference between 
Eq. \eqref{eq: nontdrl} and Eq. \eqref{eq: tsnllp} is 
\begin{equation}
   \Delta t_\text{nl} = \frac{19}{3} \Gamma_\nu^{-1} (P_m^{-\frac{1}{2}} -1), ~~ P_m<1.
\end{equation}
If the initial seed field is sufficiently weak to satisfy $\mathcal{E}_0<0.04 E_{k,\nu}$,
the ratio between the first term of $\Delta t_\text{kin}$ and $\Delta t_\text{nl}$ 
\begin{equation}
   \frac{4}{19} \ln \Big(\frac{5E_{k,\nu}}{\mathcal{E}_0}\Big)
\end{equation}
exceeds $1$, so that we have $\Delta t_\text{kin}>\Delta t_\text{nl}$, implying the low-$P_m$ turbulent 
dynamo has a longer overall timescale. 
The total time difference of the entire dynamo process is
\begin{equation}
\begin{aligned}
   \Delta t_\text{tot}  = &\Delta t_\text{kin}-\Delta t_\text{nl} \\
                               = & (P_m^{-\frac{1}{2}}-1) \Big[\frac{4}{3} \Gamma_\nu^{-1} \ln \Big(\frac{5E_{k,\nu}}{\mathcal{E}_0}\Big)-\frac{19}{3} \Gamma_\nu^{-1}\Big] \\
                                  &+\frac{4}{3} P_m^{-\frac{1}{2}} \Gamma_\nu^{-1} \ln (P_m^{-\frac{1}{2}}), ~~ P_m<1, 
\end{aligned}
\end{equation}
which increases with a decreasing $P_m$. 

The $P_m$ dependency can be clearly seen for the dynamo timescales in a conducting fluid at $P_m \leq 1$.
The overall efficiency of the dynamo increases with $P_m$, along with the weakening of magnetic energy dissipation. 
However, when $P_m$ exceeds unity, 
the dissipation effect is irrelevant to the total timescale, which becomes independent of $P_m$.

We should also point out that although the dynamo at $P_m=1$ has the same timescales for both kinematic and nonlinear stages 
as the large-$P_m$ case, 
it lacks all the evolutionary stages taking place in the sub-viscous range since
magnetic fluctuations can only survive within the undamped inertial range. 
Based on this consideration, we classify the $P_m =1 $ dynamo as low-$P_m$ case.
They both possess the distinctive damping stage, 
and both lack the $k^{-1}$ subrange of magnetic spectrum in the sub-viscous range.

\begin{figure*}[htbp]
\centering
\subfigure[$P_m=1$, this work]{
   \includegraphics[width=8cm]{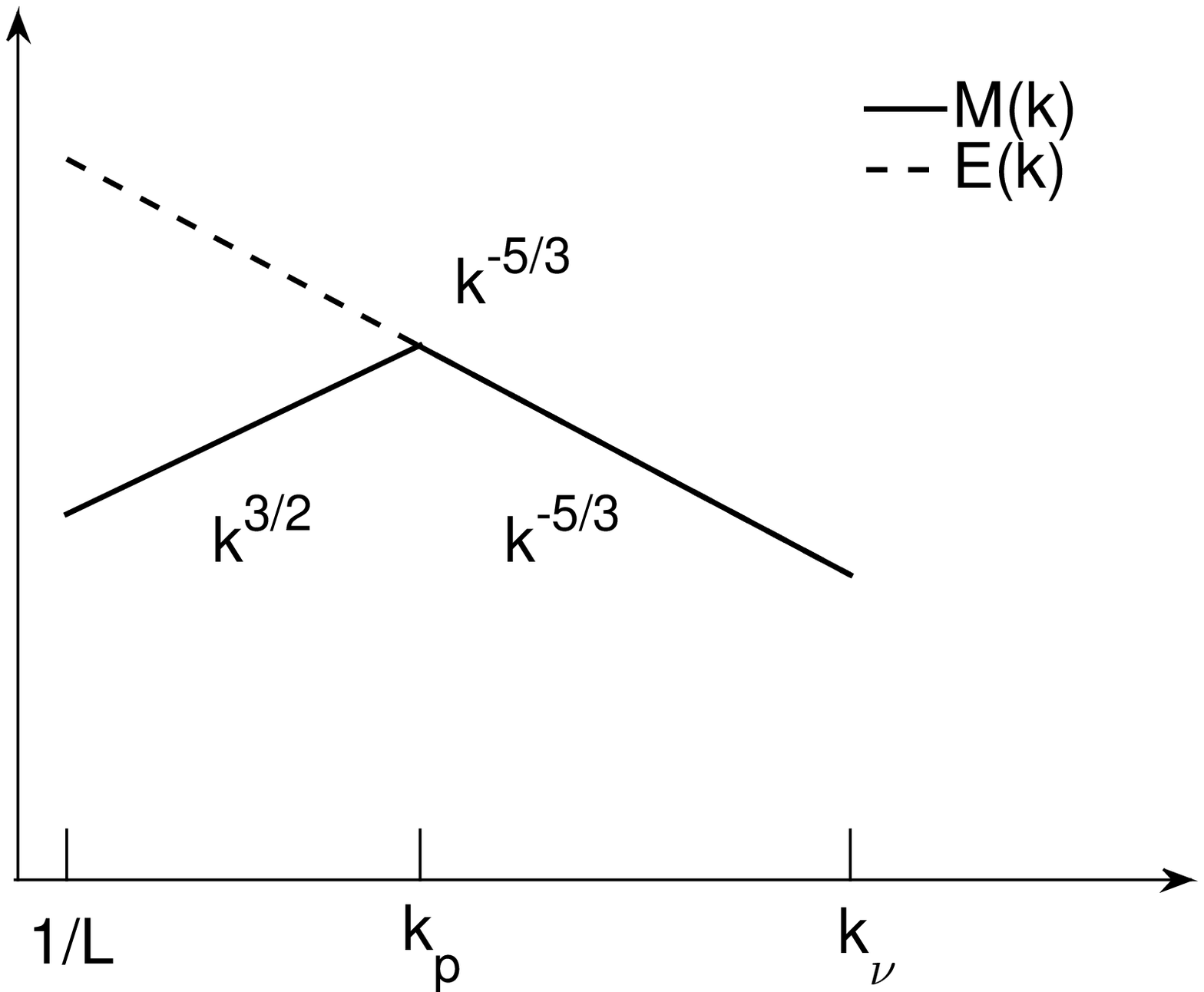}\label{fig: pmlt}}
\subfigure[$P_m>1$, this work]{
   \includegraphics[width=8cm]{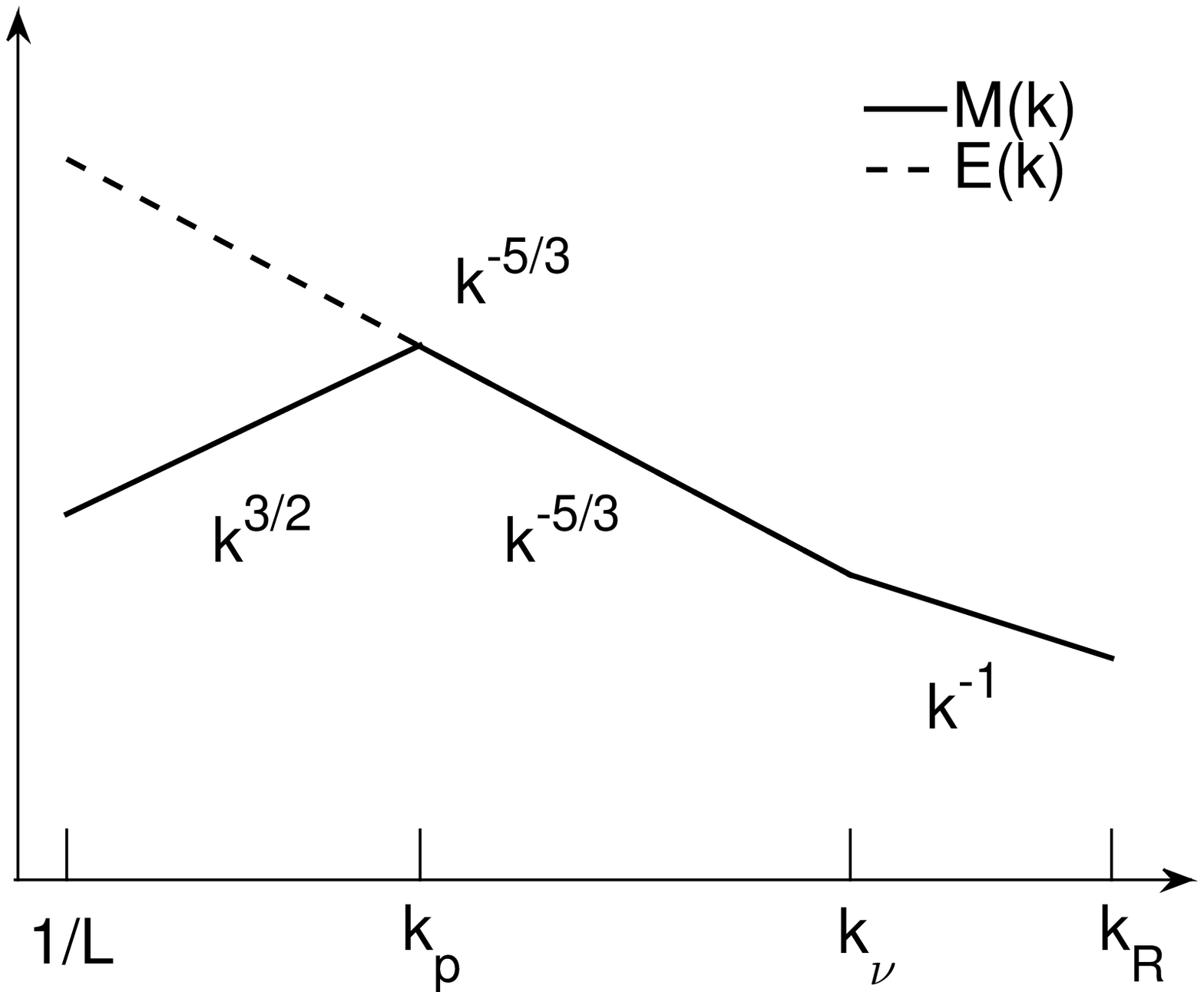}\label{fig: pmht}}
\subfigure[$P_m=1$, figure. 5.1 in \citet{Bran05}]{
   \includegraphics[width=8cm]{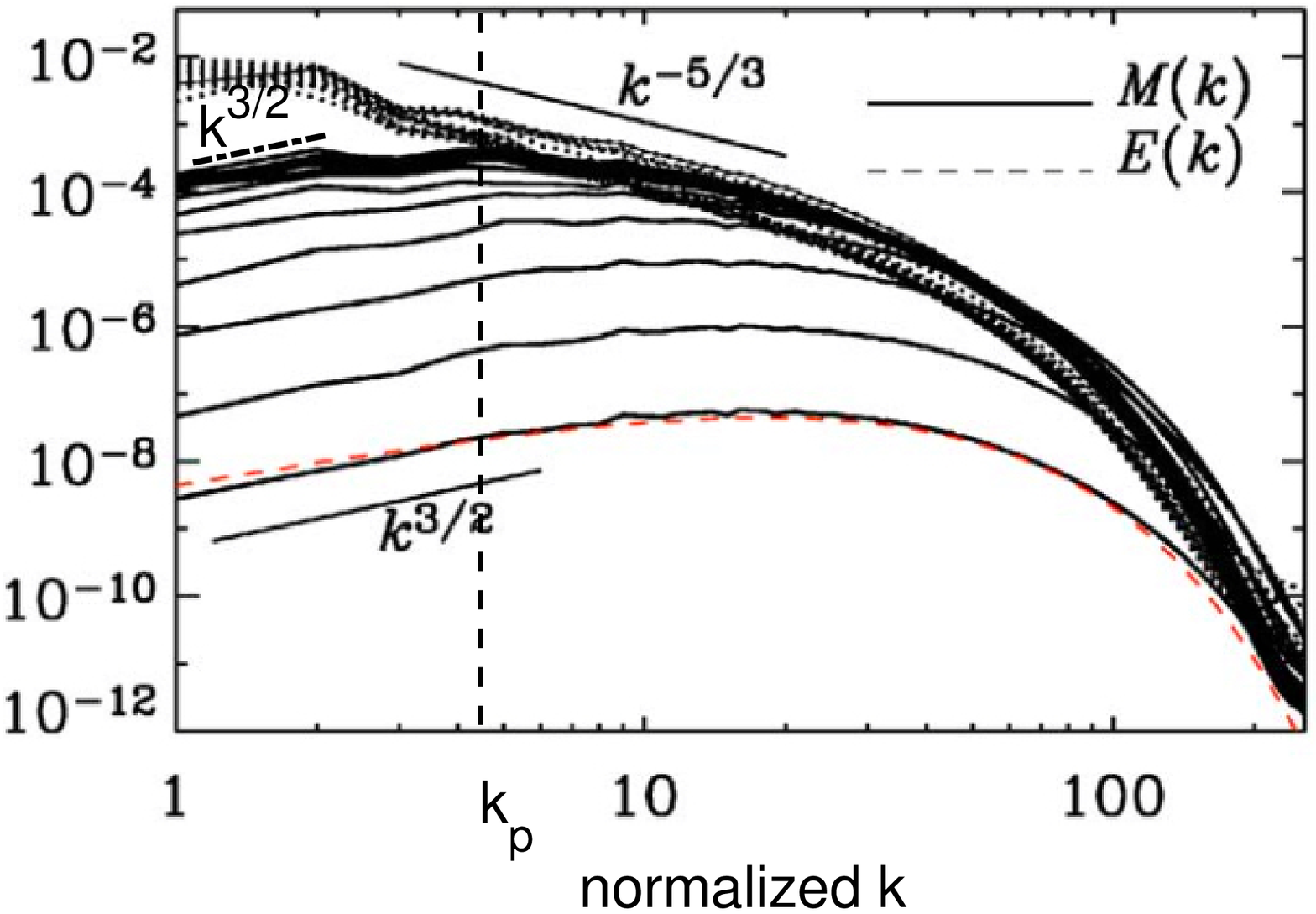}\label{fig: pmls}}
\subfigure[$P_m=50$, figure. 5.2 in \citet{Bran05}]{
   \includegraphics[width=8cm]{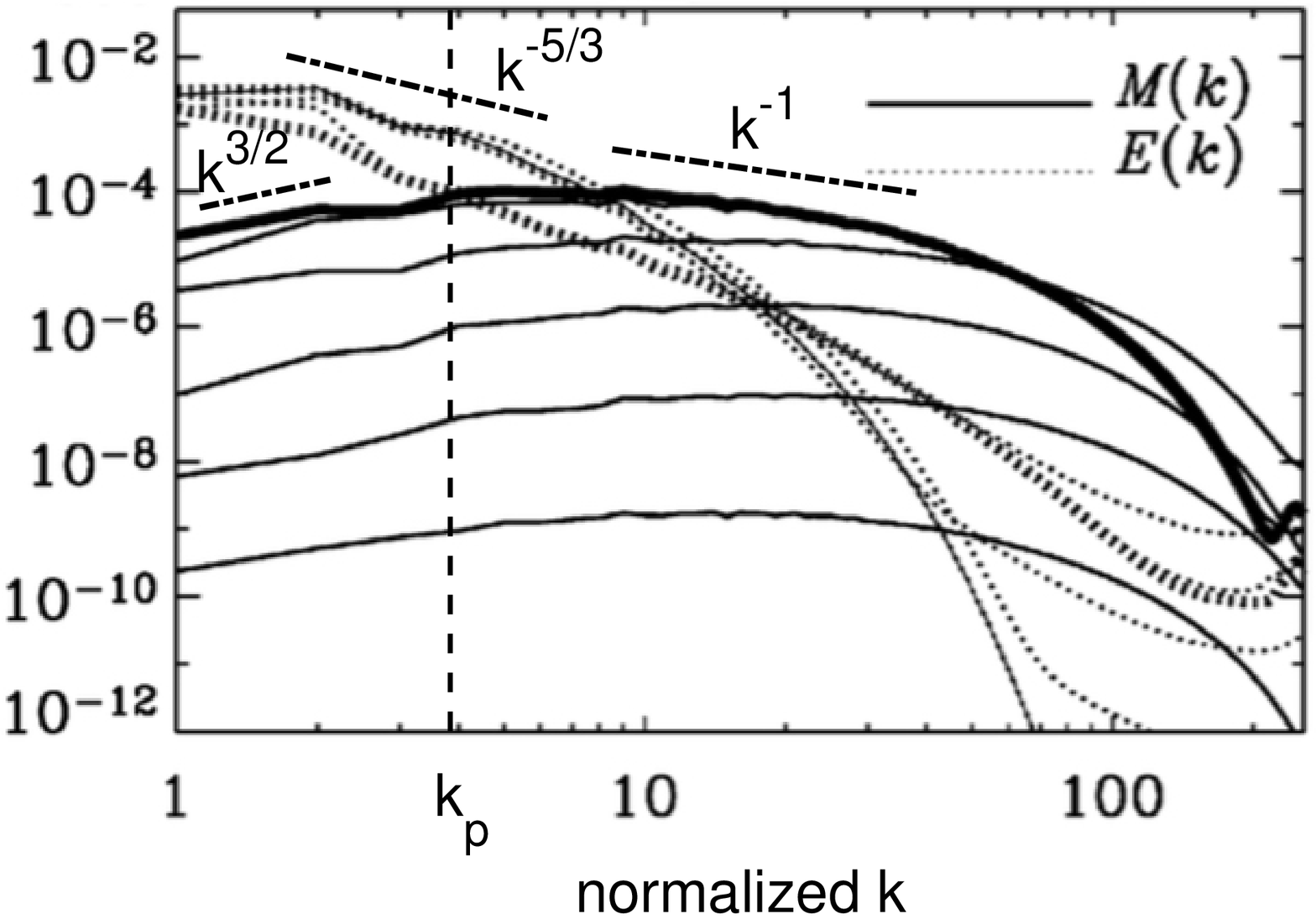}\label{fig: pmhs}}   
\caption{ Upper panel: sketches of the magnetic (solid line) and turbulent kinetic (dashed line) energy spectra in the nonlinear stage of turbulent dynamo for 
(a) $P_m =1$ and (b) $P_m>1$. 
Lower panel: (c) and (d) are figure 5.1 and figure 5.2 taken from 
\citet{Bran05} for $P_m =1$ and $P_m = 50$, respectively, where 
we add dash-dotted lines to indicate different spectral slopes and the vertical dashed line to represent the equipartition scale at the end of their 
simulations. }
\label{fig: sket}
\end{figure*}

In Figure \ref{fig: sket} we illustrate the magnetic energy spectrum in the nonlinear stage of turbulent dynamo. 
At $P_m=1$ (Fig. \ref{fig: pmlt}), it follows the Kazantsev $k^{3/2}$ profile on scales larger than $1/k_p$, while on smaller scales the transition to 
MHD turbulence occurs and there is a $k^{-5/3}$ range for both the kinetic and magnetic energies.
This theoretical expectation is consistent with the earlier numerical result from 
\citet{Bran05}
at $P_m=1$, shown in Fig. \ref{fig: pmls}.
Different solid lines represent different times for the evolution of the magnetic spectrum. 
It is obvious that initially only the Kazantsev spectrum is present until reaching larger $k$ where 
the numerical dissipation effect takes over. This corresponds to the kinematic stage. 
At later times the numerical result testifies that the nonlinear effect becomes important when 
the magnetic energy grows to equipartition with the kinetic energy. As a result, both the Kazantsev $k^{3/2}$ spectrum at $k<k_p$ and 
MHD $k^{-5/3}$ spectrum at $k>k_p$ can be seen, which agrees well with our analysis. 
The overplotted $k_p$ (vertical dashed line) denotes the equipartition scale at the end of their simulations.

We display the magnetic spectrum for $P_m>1$ dynamo in the nonlinear stage in Fig. \ref{fig: pmht}, besides the $k^{3/2}$ spectrum and 
$k^{-5/3}$ spectrum in the inertial range, which also features the $k^{-1}$ spectrum in the sub-viscous range. 
The numerical testing of $P_m>1$ dynamo is more challenging than the case of $P_m=1$
since it requires much higher numerical resolution to cover both turbulence inertial range and viscosity-dominated range, 
i.e., both large kinetic Reynolds number and large $P_m$. 
From the numerical result for $P_m=50$ shown in Fig. \ref{fig: pmhs}, we see that the kinetic energy spectrum decays with time, and in the spectrum 
for the last time the inertial range characterized by the $k^{-5/3}$ scaling is essentially absent. 
Accordingly, as a numerical artifact of insufficient inertial range, the $k^{-5/3}$ subrange is also missing in the magnetic spectrum, 
and only the $k^{3/2}$ spectrum remains at large scales. 
But in the dissipative subrange of the kinetic spectrum, the magnetic spectrum is compatible with the $k^{-1}$ slope which we analytically derived 
in the sub-viscous range. 
A better defined $k^{-1}$ range was observed in the simulations in 
\citet{Hau04}.
Hopefully, future high resolution simulations can better determine the spectral form of the magnetic energy and 
test the existence of all the predicted asymptotic slopes.

\section{Turbulent dynamo in partially ionized gas}\label{sec: partiond}

In a partially ionized medium, ions are subject to Lorentz force and tied to magnetic field lines, 
whereas neutrals are not directly affected by magnetic field. 
Due to the relative drift between the two species, neutrals exert collisional damping on the motions of ions and cause dissipation of the magnetic energy. 
Since ion-neutral collisional damping is the dominant damping process in partially ionized media like molecular clouds
\citep{KulA92}, 
we disregard the resistive damping in this case for analytical simplicity. 
The ion-neutral collisional damping rate is a function of $\mathcal{E}$
\citep{Kulsrud_Pearce, KulA92}, 
\begin{equation}\label{eq: damrt}
      \omega_d \approx \frac{\xi_n V_A^2 k_\| ^2}{2\nu_{ni}} = \frac{\xi_n V_A^2 k ^2}{6\nu_{ni}} = \mathcal{C} k^2 \mathcal{E}, 
\end{equation}
where $\xi_n=\rho_n/\rho$ is neutral fraction, $\nu_{ni}=\gamma_d \rho_i$ is neutral-ion collision frequency, 
and $\gamma_d$ is the drag coefficient introduced in 
\citet{Shu92}.
The parameter $\mathcal{C}$ is defined as $\xi_n / (3\nu_{ni})$, proportional to neutral-ion collisional time. 
We next look into the growth of magnetic energy with different ranges of ionization fractions.

\subsection{Low ionization fraction}\label{sec: parlow}

Compared with the case of fully ionized gas, the dependence of damping on magnetic energy (see Eq. \eqref{eq: damrt}) 
introduces further complications. 
Initially, when the small-scale magnetic field is sufficiently weak,  
the ion-neutral collisional damping is negligible. The kinematic dynamo operates in the dissipation-free stage.

(1) {\it Dissipation-free stage}

Following the Kazantsev theory in the kinematic regime, 
the magnetic energy grows exponentially (Eq. \eqref{eq: enedif}) and 
the spectral peak shifts to smaller scales.
The growth rate $\Gamma_\nu$ is given by Eq. \eqref{eq: grnu}, where 
$k_\nu$ is the hydrodynamic viscous scale of neutrals, 
\begin{equation}\label{eq: knu}
 k_\nu=L^{-\frac{1}{4}}V_L^{\frac{3}{4}} \nu_n^{-\frac{3}{4}}. 
\end{equation}
Here $\nu_n=c_{sn}/(n_n\sigma_{nn})$ is the kinematic viscosity, $n_n$ is neutral number density, 
$c_{sn}$ is the sound speed in neutrals, and $\sigma_{nn}$ is the cross section for a neutral-neutral collision.

On the other hand, although the turbulent diffusion of magnetic fields is not involved in the kinematic stage, the effect of 
ion-neutral collisional damping becomes more and more important with the growth of magnetic energy. 
The damping scale of magnetic fluctuations is determined by the balance between the damping rate $\omega_d$ 
and growth rate $\Gamma_\nu$. 
Combining Eq. \eqref{eq: grnu} and \eqref{eq: damrt}, we get its functional dependence on magnetic energy, 
\begin{equation}\label{eq: kddif}
 k_d  = \mathcal{C}^{-\frac{1}{2}} \Gamma_\nu^\frac{1}{2} \mathcal{E} ^{-\frac{1}{2}}
        = \mathcal{C} ^{-\frac{1}{2}} L^{-\frac{1}{6}} V_L^\frac{1}{2} k_\nu^\frac{1}{3}  \mathcal{E} ^{-\frac{1}{2}} .
\end{equation}
The damping scale increases with the growth of magnetic energy.  
Having $\mathcal{E}$ given by Eq. \eqref{eq: enedif}, it becomes 
\begin{equation}\label{eq: kddift}
    k_d  =  \mathcal{C} ^{-\frac{1}{2}} \Gamma_\nu^\frac{1}{2}  \mathcal{E}_0 ^{-\frac{1}{2}} \exp{( - \Gamma_\nu t)}.
\end{equation}
In the damped region beyond $k_d$, the slippage between neutrals and ions is significant enough to 
dissipate any magnetic fluctuation before it is built up. 
When $k_d$ approaches the peak scale of magnetic energy spectrum, over smaller scales below the spectral peak, 
the damping effect becomes significant and the dissipation-free approximation breaks down. 
Then the evolving magnetic energy undergoes the next viscous stage.

(2) {\it Viscous stage}

The magnetic energy spectrum in the viscous stage is peaked and cut off at $k_d$. 
So the magnetic energy is recast as 
\begin{equation}
\begin{aligned}
     \mathcal{E}   
     &= \frac{1}{2} \int_0 ^ {k_d} M_0 \exp{\bigg(\frac{3}{4} \Gamma_\nu t \bigg)} \bigg( \frac{k}{k_\nu} \bigg)^\frac{3}{2} dk \\
                           &=\frac{1}{5} \mathcal{E}_0 \exp{\bigg(\frac{3}{4} \Gamma_\nu t \bigg)}  \Big(\frac{k_d}{k_\nu} \Big)^\frac{5}{2}.
\end{aligned}
\end{equation}
Inserting Eq. \eqref{eq: kddif} in the above equation, we obtain 
\begin{equation}\label{eq: enedamr}
    \mathcal{E}= \Big(\frac{1}{5}\Big)^\frac{4}{9} \Big(\frac{\Gamma_\nu}{\omega_{d0,\nu}}\Big)^\frac{5}{9}  \mathcal{E}_0 \exp{\Big(\frac{1}{3}\Gamma_\nu t\Big)},
\end{equation}
and 
\begin{equation}\label{eq: dagdar}
   k_d= 5^\frac{2}{9} \Big(\frac{\Gamma_\nu}{\omega_{d0,\nu}}\Big)^\frac{2}{9} k_\nu \exp{\Big(-\frac{1}{6}\Gamma_\nu t\Big)},
\end{equation}
where $ \omega_{d0,\nu}= \mathcal{C} k_\nu^2 \mathcal{E}_0$ is the initial ion-neutral collisional damping rate at $k_\nu$. 
The modified dynamo growth with a reduced exponential growth rate ($\sim e^{\Gamma_\nu t/3}$) was also derived by 
\citet{KulA92}
in their studies of turbulent dynamo in partially ionized gas. 
The above expressions imply that a more effective neutral-ion coupling with a higher $\nu_{ni}$ (smaller $\mathcal{C}$) leads to 
a higher magnetic energy and smaller damping scale in the viscous stage.

The crossing time between the dissipation-free and viscous stages can be given by equalizing
$\mathcal{E}$ in Eq. \eqref{eq: enedif} with $\mathcal{E}$ in Eq. \eqref{eq: enedamr},
\begin{equation}\label{eq: tim12}
   t_{12} =  \Gamma_\nu^{-1} \ln \Big[\Big(\frac{1}{5}\Big)^\frac{4}{15} \Big(\frac{\Gamma_\nu}{\omega_{d0,\nu}}\Big)^\frac{1}{3} \Big].
\end{equation}
The magnetic energy and damping scale at $t_{12}$ are 
\begin{equation}\label{eq: ene12}
    \mathcal{E}(t_{12}) =  \Big(\frac{1}{5}\Big)^\frac{8}{15} \Big(\frac{\Gamma_\nu}{\omega_{d0,\nu}}\Big)^\frac{2}{3} \mathcal{E}_0, 
\end{equation}
and 
\begin{equation}\label{eq: kd12}
  k_d(t_{12}) = 5^\frac{4}{15} \Big(\frac{\Gamma_\nu}{\omega_{d0,\nu}}\Big)^\frac{1}{6} k_\nu.
\end{equation}

The viscous stage proceeds until the damping scale approaches the viscous scale, i.e., $k_p=k_d=k_\nu$. 
The corresponding time can be computed by equaling $k_d$ (Eq. \eqref{eq: dagdar}) with $k_\nu$, 
\begin{equation}\label{eq: t23}
   t_{23}= \frac{4}{3} \Gamma_\nu^{-1} \ln \Big[5 \Big(\frac{\Gamma_\nu}{\omega_{d0,\nu}}\Big)  \Big] . 
\end{equation}
The duration for the viscous stage can then be determined (Eq. \eqref{eq: tim12} and \eqref{eq: t23}), 
\begin{equation}\label{eq: tranvis}
   \tau_\text{vis} = t_{23}-t_{12}= \Gamma_\nu^{-1} \ln \Big[5^\frac{8}{5} \Big(\frac{\Gamma_\nu}{\omega_{d0,\nu}}\Big)\Big]. 
\end{equation}
We find that the magnetic energy reached at $t_{23}$ is (Eq. \eqref{eq: enedamr} and \eqref{eq: t23})
\begin{equation}\label{eq: ene23}
    \mathcal{E}(t_{23})= \Big(\frac{\Gamma_\nu}{\omega_{d0,\nu}}\Big) \mathcal{E}_0,
\end{equation}
which can also be equivalently obtained from the relation $\Gamma_\nu=\omega_d$ at $k_\nu$. 
The ratio between $\mathcal{E}(t_{23})$ and the kinetic energy at $k_\nu$ (Eq. \eqref{eq: kinenu}) is 
\begin{equation}\label{eq: ratgam}
  \mathcal{R} = \frac{\mathcal{E}(t_{23})}{E_{k,\nu}} =\frac{2\mathcal{C}^{-1}}{\Gamma_\nu} = \frac{6}{\xi_n} \frac{\nu_{ni}}{\Gamma_\nu}. 
\end{equation}
It relates to the ratio $(\Gamma_\nu/\omega_{d0,\nu})$ by (Eq. \eqref{eq: ene23}, \eqref{eq: ratgam})
\begin{equation}\label{eq: relapr}
    \frac{\Gamma_\nu}{\omega_{d0,\nu}}= \mathcal{R} \frac{E_{k,\nu}}{\mathcal{E}_0}.
\end{equation}

The parameter $\mathcal{R}$ can be viewed as an indicator of the degree of ionization $\xi_i= \rho_i/\rho$, 
which determines the coupling degree of neutrals with ions.  
At the beginning stage of the dynamo, the magnetic field is too weak to manifest itself, so neutrals and ions 
can be treated together as a single fluid. The growth of magnetic energy is driven by the hydrodynamic turbulent motions of both 
neutrals and ions.
Meanwhile, the increase of magnetic field strength gives rise to relative drift between the two species. That induces the 
collisional dissipation of magnetic energy. 
The condition $\mathcal{R}=1$ corresponds to a critical ionization fraction 
\begin{equation}\label{eq: crionfrc}
        \xi_{i,\text{cr}} = \frac{\Gamma_\nu}{6 \gamma_d \rho + \Gamma_\nu} 
        = \frac{L^{-\frac{1}{2}} V_L^\frac{3}{2} \nu_n^{-\frac{1}{2}}}{6 \gamma_d \rho + L^{-\frac{1}{2}} V_L^\frac{3}{2} \nu_n^{-\frac{1}{2}}}.
\end{equation}
When $\mathcal{R}<1$, namely, $\xi_i<\xi_{i,\text{cr}}$, neutral-ion collisions are not frequent enough to ensure a strong coupling, 
but instead the damping effect is enhanced. 
We can also see from Eq. \eqref{eq: ratgam} that at $\mathcal{R}<1$, there exists $\nu_{ni}<\Gamma_\nu$. 
It means the neutral-ion collisions are inefficient in converting the kinetic energy carried by the viscous-scale eddies 
to magnetic energy. 
As a result, the magnetic energy accumulated at the end of the viscous stage is still unsaturated with 
$\mathcal{E}(t_{23})<E_{k,\nu}$, whereupon the kinematic stage proceeds to the inertial range of turbulence.
We first deal with the case of $R<1$ in weakly ionized medium, and then turn to the case of $R \geq 1$ at a high ionization fraction in 
Section \ref{eq: sechif}.

(3) {\it Damping stage}

When the damping scale arrives at the viscous scale, the equalization between the growth rate and dissipation rate of magnetic 
energy terminates the dynamo growth at the viscous scale. 
Hence the peak of magnetic energy spectrum shifts to a somewhat larger scale 
where the eddy turnover rate exceeds the damping rate, until the new equilibrium between the eddy turnover rate and damping rate is reestablished. 
The spectral peak moves to ever-larger scales, following which the damping scale keeps increasing.
The turbulent eddies below the damping scale have their turnover rates smaller than the damping rate and thus are unable to amplify the magnetic field. 
The turbulent eddies at the damping scale are responsible for the dynamo growth. 
Since larger-scale eddies have slower turnover rates,  
the corresponding growth rate decreases with the increase of the magnetic energy, as well as the damping scale, 
\begin{equation}\label{eq: gamlak}
   \Gamma \sim \Gamma_d= L^{-\frac{1}{3}}V_Lk_d^\frac{2}{3}.
\end{equation}
Together with Eq. \eqref{eq: damrt}, \eqref{eq: gamlak}, the balance $\Gamma_d=\omega_d$ yields
\begin{equation}\label{eq: kdlak}
    k_d=\mathcal{C}^{-\frac{3}{4}} L^{-\frac{1}{4}} V_L^\frac{3}{4} \mathcal{E}^{-\frac{3}{4}}.
\end{equation}
By describing the magnetic energy as
\begin{equation}\label{eq: enlak}
\begin{aligned}
  \mathcal{E}
  & = \frac{1}{2} \int_0 ^ {k_d} M_0 \exp{\bigg(\frac{3}{4} \int \Gamma dt \bigg)} \bigg( \frac{k}{k_\nu} \bigg)^\frac{3}{2} dk \\
                     & = \frac{1}{5} \mathcal{E}_0  \exp{\bigg(\frac{3}{4} \int \Gamma dt \bigg)} \Big(\frac{k_d}{k_\nu}\Big)^\frac{5}{2},
\end{aligned}
\end{equation}
we can obtain 
\begin{equation}
    \frac{d \ln \mathcal{E}}{ dt} = \frac{3}{4} \Gamma + \frac{5}{2} \frac{d \ln k_d}{dt}.
\end{equation}
By using the relation in Eq. \eqref{eq: kdlak}, the above equation leads to 
\begin{equation}\label{eq: engenk}
    \frac{d \mathcal{E}}{\mathcal{E}} = \frac{6}{23} \Gamma dt . 
\end{equation}
Combining Eq. \eqref{eq: gamlak}, \eqref{eq: kdlak}, and \eqref{eq: engenk}, we derive
\begin{equation}
      \sqrt{\mathcal{E}} \sim \frac{3}{23} \mathcal{C}^{-\frac{1}{2}} L^{-\frac{1}{2}} V_L^\frac{3}{2} t, 
\end{equation}
which shows that the magnetic field strength grows proportionally to time, i.e., $\mathcal{E} \sim t^2$, in the damping stage.
While the linear dependence of magnetic energy on time in the nonlinear stage of dynamo has been observed in MHD simulations 
(e.g., \citealt{CVB09, Bere11}),
this quadratic dependence of magnetic energy on time in the damping stage identified in our analysis should be tested by future two-fluid simulations.
Using the energy and time at the onset of damping stage as the boundary condition of Eq. \eqref{eq: engenk}, 
we can further get the exact expression of magnetic energy, 
\begin{equation}\label{eq: enetev3}
      \sqrt{\mathcal{E}} = \sqrt{\mathcal{E}(t_{23})} + \frac{3}{23} \mathcal{C}^{-\frac{1}{2}} L^{-\frac{1}{2}} V_L^\frac{3}{2} (t-t_{23}).
\end{equation}
Its insertion in Eq. \eqref{eq: kdlak} yields 
\begin{equation}\label{eq: kddarap}
    k_d = \Big[k_\nu^{-\frac{2}{3}}+\frac{3}{23}L^{-\frac{1}{3}}V_L (t-t_{23})\Big]^{-\frac{3}{2}}.
\end{equation}

The damping stage persists until the magnetic energy is built up to reach equilibrium with the kinetic energy,
\begin{equation}\label{eq: enevel}
\mathcal{E}=\frac{1}{2} v_d^2 = \frac{1}{2} L^{-\frac{2}{3}} V_L^2 k_d^{-\frac{2}{3}} , 
\end{equation}
at the critical damping scale (Eq. \eqref{eq: kdlak} and \eqref{eq: enevel}),
\begin{equation}\label{eq: cridams}
   k_{d,\text{cr}} = \Big(\frac{\mathcal{C}}{2}\Big)^{-\frac{3}{2}} L^\frac{1}{2} V_L^{-\frac{3}{2}} = \mathcal{R}^\frac{3}{2} k_\nu. 
\end{equation}
The kinematic saturation eventually occurs at a scale $\mathcal{R}^{-3/2}$ times larger than the viscous scale. 
The corresponding critical energy can be found by substituting the above expression in Eq. \eqref{eq: enevel}, 
\begin{equation} \label{eq: criene}
   \mathcal{E}_\text{cr}= \frac{1}{2} L^{-\frac{2}{3}} V_L^2 k_{d,\text{cr}}^{-\frac{2}{3}} = \frac{\mathcal{C}}{4} L^{-1} V_L^3 = \mathcal{R}^{-1}E_{k,\nu} ,
\end{equation}
where the relation in Eq. \eqref{eq: ratgam} is used.
It is inversely proportional to $ \mathcal{R}$ and thus has a larger value at a low degree of ionization 
when ion-neutral collisional damping is strong. 
Once this level of magnetic energy is achieved, the nonlinear stage is initiated. 
The time for the onset of nonlinearity can be determined by inserting $\mathcal{E}=\mathcal{E}_\text{cr}$ to Eq. \eqref{eq: enetev3}. 
That is 
\begin{equation}\label{eq: timsct}
   t_\text{cr}= \frac{23}{3} \Big( \frac{\mathcal{C}}{2} - \Gamma_\nu^{-1} \Big) + t_{23}=\frac{23}{3} \Gamma_\nu^{-1} (\mathcal{R}^{-1}-1)+t_{23}.      
\end{equation}
So the damping stage lasts for  
\begin{equation}\label{eq: tiscdr}
    \tau_\text{dam}= \frac{23}{3} \Gamma_\nu^{-1} (\mathcal{R}^{-1}-1).
\end{equation}
Apparently, a sufficiently small $\mathcal{R}$ in weakly ionized medium
can lead to an extended growth history of magnetic energy in the damping stage.

In contrast to the exponential growth of magnetic energy in the dissipation-free and viscous stages,
the magnetic energy in the damping stage
is not only largely consumed by more severe collisional damping, but less efficiently amplified by eddies with slower turnover 
rates than the viscous-scale eddies in accordance with the increasing damping scale. 
This results in a linear growth of magnetic field strength in time and slower approach to the kinematic saturation.

(4) {\it Nonlinear stage}

The nonlinear stage adheres to the evolution law formulated in Eq. \eqref{eq: ennoncr}. 
At the final equipartition state, the growth time is (Eq. \eqref{eq: ennoncr}, \eqref{eq: satnlf}, and \eqref{eq: criene})
\begin{equation}\label{eq: t4}
   t_4=\frac{19}{3} \Big(\frac{L}{V_L}-\frac{\mathcal{C}}{2}\Big) + t_\text{cr} = \frac{19}{3} \Big(\frac{L}{V_L}-\mathcal{R}^{-1} \Gamma_\nu^{-1}\Big) + t_\text{cr} .
\end{equation}
From that we see the duration of the nonlinear stage is 
\begin{equation}\label{eq: tscnli}
   \tau_\text{nl}= \frac{19}{3} \Big(\frac{L}{V_L}-\mathcal{R}^{-1} \Gamma_\nu^{-1}\Big) .
\end{equation}
In the case of weakly ionized gas with a small $\mathcal{R}$, 
the kinematic stage can bring the magnetic energy to a level appreciably higher 
than the turbulent energy of the smallest eddies. Therefore the following nonlinear stage is accordingly shortened. 
We further substitute Eq. \eqref{eq: t23}, \eqref{eq: timsct} into Eq. \eqref{eq: t4} and use Eq. \eqref{eq: relapr}
to write the full expression of the final saturation time, 
\begin{equation}\label{eq: rst4}
   t_4= \frac{19}{3} \frac{L}{V_L} +\frac{4}{3} \mathcal{R}^{-1} \Gamma_\nu^{-1} - \frac{23}{3} \Gamma_\nu^{-1} + 
            \frac{4}{3} \Gamma_\nu^{-1} \ln \Big[5 \mathcal{R} \Big(\frac{E_{k,\nu}}{\mathcal{E}_0}\Big)  \Big] .
\end{equation}
Presumably, the first term dominantly contributes to the total timescale, thus during the whole turbulent dynamo process
the largest eddy turns over around 6 times.

From the comparison between the kinematic and nonlinear stages of the turbulent dynamo, we find that 
in the presence of ion-neutral collisional damping, the kinematic stage is strongly modified, with the magnetic energy 
having a reduced exponential growth rate in its viscous stage, and a quadratic growth rate in its damping stage. 
In contrast, the magnetic energy growth during the nonlinear stage is unaffected by the energy dissipation process. 
Its linear dependence on time is simply determined by the properties of MHD turbulence.

\subsection{High ionization fraction ($R \geq 1$)}\label{eq: sechif}

In the kinematic stage, when the magnetic energy approaches equipartition with the turbulent energy contained in the viscous-scale 
eddies, substituting $\mathcal{E}=E_{k,\nu}$ (Eq. \eqref{eq: kinenu}) in Eq. \eqref{eq: kddif} leads to 
\begin{equation}\label{eq: kdknu}
    k_d = \Big(\frac{\mathcal{C}}{2}\Big)^{-\frac{1}{2}} L^\frac{1}{6} V_L^{-\frac{1}{2}} k_\nu^\frac{2}{3} = \sqrt{\mathcal{R}} ~k_\nu. 
\end{equation}
Given this expression, $k_d < k_\nu$ is equivalent to $\mathcal{R}<1$, corresponding to the situations discussed above. 
We next turn to other situations where $k_d \geq k_\nu$ is satisfied, with $\mathcal{R}\geq1$ (i.e. $\xi_i \geq \xi_{i,\text{cr}}$).

In the case of $\mathcal{R} =1$, the growth of magnetic energy undergoes the same 
dissipation-free and viscous stages as described in Section \ref{sec: parlow}. 
Once the kinematic dynamo saturates at $k_\nu$ in the viscous stage, the damping stage no longer exists with 
$\tau_\text{dam}=0$ (Eq. \eqref{eq: tiscdr}), 
and the nonlinear stage commences.
The universal treatment of the nonlinear stage applies. 
As the condition for the onset of nonlinearity, 
the critical energy is $\mathcal{E}_\text{cr} = E_{k,\nu}$ (Eq. \eqref{eq: criene}), and 
the critical time $t_\text{cr}=t_{23}$ (Eq. \eqref{eq: timsct}) coincides with the critical time in the case of a conducting fluid at $P_m>1$ 
(Eq. \eqref{eq: tnlhesv}). 
The time interval $\tau_{nl}$ in Eq. \eqref{eq: tscnli} now takes the same form as in Eq. \eqref{eq: nontdrl}.

In the case of $\mathcal{R} >1$, we first consider the scenario where the kinematic dynamo saturates in the dissipation-free stage. 
The magnetic energy initially grows according to Eq. \eqref{eq: enedif}, until saturates at the time $t_\text{sat,k}$, which is expressed 
as in Eq. \eqref{eq: dfrtcr}. 
To accommodate this possibility, it requires $t_\text{sat,k}$ is shorter than $t_{12}$ from Eq. \eqref{eq: tim12}. This confines 
\begin{equation}
   \mathcal{R} \geq 5^\frac{4}{5} \Big(\frac{E_{k,\nu}}{\mathcal{E}_0}\Big)^\frac{1}{2}.
\end{equation}
At a lower $ \mathcal{R}$, the viscous stage is present. 
The equalization between $\mathcal{E}$ in Eq. \eqref{eq: enedamr} and $E_{k,\nu}$ gives the saturation time for the kinematic dynamo,
\begin{equation}\label{eq: tcrrla}
\begin{aligned}
    t_\text{sat,k}&= \frac{4}{3} \Gamma_\nu^{-1} \ln \Big[ 5 \mathcal{R}^{-\frac{9}{4}} \Big(\frac{\Gamma_\nu}{\omega_{d0,\nu}}\Big)\Big]   \\
                   &=\frac{4}{3} \Gamma_\nu^{-1} \ln \Big[ 5 \mathcal{R}^{-\frac{5}{4}} \Big(\frac{E_{k,\nu}}{\mathcal{E}_0}\Big)\Big].   
\end{aligned}
\end{equation}
The damping scale at $t_\text{sat,k}$ calculated from Eq. \eqref{eq: dagdar}, \eqref{eq: tcrrla}
is consistent with the result presented in Eq. \eqref{eq: kdknu},
\begin{equation}\label{eq: hikdve}
    k_d (t_\text{sat,k}) =\sqrt{ \mathcal{R} } k_\nu. 
\end{equation}
Therefore the viscous stage has a time span (Eq. \eqref{eq: tim12} and \eqref{eq: tcrrla})
\begin{equation}\label{eq: rltvis}
\begin{aligned}
   \tau_\text{vis} = t_\text{sat,k}-t_{12}&= \Gamma_\nu^{-1} \ln \Big[5^{\frac{8}{5}} \mathcal{R}^{-3} \Big(\frac{\Gamma_\nu}{\omega_{d0,\nu}}\Big)  \Big] \\
   &=  \Gamma_\nu^{-1} \ln \Big[5^{\frac{8}{5}}  \mathcal{R}^{-2} \Big(\frac{E_{k,\nu}}{\mathcal{E}_0}\Big) \Big],
\end{aligned}
\end{equation}
which has a term $\mathcal{R}^{-3}$ in the logarithm and thus is shorter than the $\tau_\text{vis}$ in the case of $\mathcal{R} <1$ (Eq. \eqref{eq: tranvis}). 

For both scenarios under the condition of $\mathcal{R} >1$, after the saturation of the kinematic dynamo at $t_\text{sat,k}$, 
the kinematic stage goes through the transitional stage, wherein the spectral peak shifts to the viscous scale and 
leaves a spectrum $\sim k^{-1}$ (Eq. \eqref{eq: negspec}) in the sub-viscous region. 
The transitional stage results in the same critical energy and critical time as in the case of $\mathcal{R} =1$, as well as 
the case of a conducting fluid at $P_m>1$. \footnote{In spite of the same formulae for $\mathcal{E}_\text{cr}$ and $t_\text{cr}$,
the viscosity involved in cases of fully and partially 
ionized gases are different.}  
It has a time interval as expressed in Eq. \eqref{eq: trantlr} for $\mathcal{R} \geq 5^{4/5} \sqrt{E_{k,\nu}/\mathcal{E}_0}$, 
and (Eq. \eqref{eq: tnlhesv}, \eqref{eq: tcrrla})
\begin{equation}\label{eq: ttranlr}
   \tau_\text{tran} =  \frac{5}{3} \Gamma_\nu^{-1} \ln \mathcal{R}
\end{equation}
for a lower $\mathcal{R}$.

During the kinematic stage, the time-evolution of magnetic energy has evident dependence on $\mathcal{R}$. On the other hand, 
in the following nonlinear stage, regardless of the range of $\mathcal{R}$, the growth of magnetic energy follows the 
universal behavior as described in Eq. \eqref{eq: ennoncr}. 
All scenarios at $\mathcal{R}\geq 1$ share the same expressions for the duration of the nonlinear stage as in Eq. \eqref{eq: nontdrl}, 
and total time of the entire dynamo process as in Eq. \eqref{eq: rltfsa}.

We are now able to compare the timescales of dynamo growth in different ranges of $\mathcal{R}$. 
It turns out that the entire timescale of the turbulent dynamo at $\mathcal{R}<1$ is longer than that at $\mathcal{R}\geq1$
by (Eq. \eqref{eq: rltfsa} and \eqref{eq: rst4})
\begin{equation}
    \Delta t_\text{tot} = t_4- t_\text{sat,nl} = \frac{4}{3} \Gamma_\nu^{-1} (\mathcal{R}^{-1}+\ln \mathcal{R} -1), 
\end{equation}
where $\mathcal{R}$ is from Eq. \eqref{eq: rltfsa} and smaller than unity. 
It increases with a decreasing $\mathcal{R}$.
Because the nonlinear stage at $\mathcal{R}\geq1$ lasts for a longer period than that at smaller $\mathcal{R}$ with the time difference 
(Eq. \eqref{eq: nontdrl} and \eqref{eq: tscnli}), 
\begin{equation}
      \Delta t_\text{nl} = \frac{19}{3} \Gamma_\nu^{-1} (\mathcal{R}^{-1} -1 ),   ~~\mathcal{R}<1, 
\end{equation}
more time is distributed to the kinematic stage at $\mathcal{R}<1$, namely, 
\begin{equation}
     \Delta t_\text{kin} = \Delta t_\text{tot} + \Delta t_\text{nl} = \frac{1}{3} \Gamma_\nu^{-1} (23 \mathcal{R}^{-1} +4 \ln \mathcal{R} -23).
\end{equation}

The above results demonstrate that in weakly ionized gas with $\xi_i< \xi_{i,\text{cr}}$, 
the magnetic field can be more efficiently amplified with the increase of ionization fraction and thus strengthening of neutral-ion coupling.  
But when the ionization is substantially enhanced with $\xi_i \geq \xi_{i,\text{cr}}$, 
the damping stage is absent, and the overall efficiency of the turbulent dynamo remains unchanged.

\subsection{Dependence of the kinematic stage on $P_m$ and $\mathcal{R}$}

Tables \ref{tab: reg1}-\ref{tab: reg5} list the evolutionary stages of magnetic energy for different ranges of $P_m$ and 
$\mathcal{R}$. The expressions of the magnetic energy $\mathcal{E}$, the peak scale $k_p$ where $\mathcal{E}$ is 
concentrated, and the corresponding time $t$ are summarized. 
By comparing the results in the cases of a conducting fluid and partially ionized gas, 
we can easily observe the close similarity in their expressions in terms of $P_m$ and $\mathcal{R}$, respectively. 
In fact, $P_m$ can also be written as the ratio of the growth rate and damping rate at $k_\nu$, 
\begin{equation}
    P_m = \frac{k_\nu^2 \nu}{k_\nu^2 \eta} = \frac{\Gamma_\nu}{k_\nu^2 \eta}. 
\end{equation}
As regards the parameter $\mathcal{R}$, from Eq. \eqref{eq: relapr} we can deduce 
\begin{equation}\label{eq: rreaex}
    \mathcal{R} = \frac{\Gamma_\nu}{k_\nu^2 E_{k,\nu} \mathcal{C}} .
\end{equation}
The term $k_\nu^2 E_{k,\nu} \mathcal{C}$ is actually the ion-neutral collisional damping rate corresponding to the kinetic energy at $k_\nu$ (see Eq. \eqref{eq: damrt}).
Thus, analogous to $P_m$, the parameter $\mathcal{R}$ can also be treated as a ratio between the growth rate and damping rate at $k_\nu$, 
but in partially ionized gas. 
Both $P_m$ and $\mathcal{R}$ are indicators of the relative importance of energy dissipation with respect to 
energy growth on the viscous scale. When their values are above unity, the saturation of the kinematic dynamo can be achieved in 
the sub-viscous range. 
Otherwise the kinematic stage extends into the inertial range, 
up to a certain scale where the local $P_m$  
\begin{equation}
    P_m (k_R) =  \frac{\Gamma_R}{k_R^2 \eta} , 
\end{equation}
or local $\mathcal{R}$
\begin{equation}
   \mathcal{R} (k_{d,\text{cr}} )  = \frac{\Gamma (k_{d,\text{cr}})}{k_{d,\text{cr}}^2 \mathcal{E}_\text{cr} \mathcal{C}} 
\end{equation}
reaches $1$.

As a natural result, the two sets of expressions formally resemble each other, except that in Table \ref{tab: reg1}, both dissipation-free and viscous stages 
are absent at $P_m \leq 1$ due to the constant resistive damping scale during the kinematic stage. 
We see that the evolution of magnetic energy at $P_m>1$ and $\mathcal{R}>1$ are very much alike and have formulae in the same pattern. 
Especially as shown in Table \ref{tab: reg5}, when the values of $P_m$ and $\mathcal{R}$ are sufficiently high to exceed 
the threshold $ 5^{4/5} \sqrt{E_{k,\nu}/\mathcal{E}_0}$, 
the dissipation process becomes irrelevant in affecting the kinematic stage 
and the expressions are independent of $P_m$ and $\mathcal{R}$.

\begin{table*}[t]
\renewcommand\arraystretch{1.7}
\centering
\begin{threeparttable}
\caption[]{}\label{tab: reg1} 
  \begin{tabular}{c|c|c|c|c|c|c|c|c}
      \toprule
     Stages                      &   \multicolumn{2}{c|}{Dissipation-free}  &     \multicolumn{2}{c|}{Viscous}    &    \multicolumn{2}{c|}{Damping}       &   \multicolumn{2}{c}{Nonlinear }          \\
     \hline
    \multicolumn{9}{c}{Conducting fluid, $P_m \leq 1$} \\
    \hline
    \multirow{2}*{ $\mathcal{E}$}
    & & & & &  $\sim e^{\frac{3}{4}\Gamma_R t}$  & $\mathcal{E}_\text{cr}$   & $\sim t$                      &  $\mathcal{E}_\text{sat,nl}$    \\
    & & & & & Eq. \eqref{eq: drlpene}          &  Eq. \eqref{eq: ekrlp}    &  Eq. \eqref{eq: ennoncr}   &  Eq. \eqref{eq: satnlf}  \\
    \hline
     \multirow{2}*{ $k_p$}
    & & & & &  $k_R$                               & $k_\text{cr}$                                   &      $\sim t^{-\frac{3}{2}}$                   &  $L^{-1}$ \\
    & & & & &  Eq. \eqref{eq: lpkrdr}        & $k_R$  (Eq. \eqref{eq: lpkrdr})        &      Eq. \eqref{eq: evokpcri}               &                 \\
    \hline
    \multirow{2}*{$t$}
    & & & & &                                              & $t_\text{cr}$                      & $\tau_\text{nl}$             & $t_2$  \\
    & & & & &                                              &  Eq. \eqref{eq: crtlp}       &    Eq. \eqref{eq: tsnllp}    & Eq. \eqref{eq: t2lp} \\
    \hline
     \multicolumn{9}{c}{Partially ionized gas, $\mathcal{R}<1$} \\
     \hline
    \multirow{2}*{ $\mathcal{E}$}
    &  $\sim e^{2\Gamma_\nu t}$ & $\mathcal{E}(t_{12})$  & $\sim e^{\frac{1}{3}\Gamma_\nu t}$ & $\mathcal{E}(t_{23})$  & $\sim t^2$                         & $\mathcal{E}_\text{cr}$ & $\sim t$                           & $\mathcal{E}_\text{sat,nl}$     \\
    &  Eq. \eqref{eq: enedif}         & Eq. \eqref{eq: ene12}  &  Eq. \eqref{eq: enedamr}          &  Eq. \eqref{eq: ene23}  & Eq. \eqref{eq: enetev3}    &  Eq. \eqref{eq: criene}  &  Eq. \eqref{eq: ennoncr}  &  Eq. \eqref{eq: satnlf}          \\
    \hline
    \multirow{2}*{ $k_p$}
& $\sim e^{\frac{3}{5} \Gamma_\nu t}$ & $k_d(t_{12})$            & $\sim e^{-\frac{1}{6}\Gamma_\nu t}$ &  $k_\nu$    & $\sim t^{-\frac{3}{2}}$          & $k_\text{cr}$              & $\sim t^{-\frac{3}{2}}$       &   $L^{-1}$ \\
& Eq. \eqref{eq: kpdif}                       & Eq. \eqref{eq: kd12}  & Eq. \eqref{eq: dagdar}                      &  Eq. \eqref{eq: knu}    &  Eq. \eqref{eq: kddarap}     & $k_{d,\text{cr}}$ (Eq. \eqref{eq: cridams})   &  Eq. \eqref{eq: evokpcri} &                  \\
    \hline
    \multirow{2}*{$t$}
    &                                            &  $t_{12}$                       &        $\tau_\text{vis}$                &  $t_{23}$                      & $\tau_\text{dam}$             & $t_\text{cr}$                 &   $\tau_\text{nl}$               &  $t_4$       \\
    &                                            &   Eq. \eqref{eq: tim12}   &      Eq. \eqref{eq: tranvis}         &  Eq. \eqref{eq: t23}       &  Eq. \eqref{eq: tiscdr}       &  Eq. \eqref{eq: timsct}  &   Eq. \eqref{eq: tscnli}       &  Eq. \eqref{eq: t4}    \\
    \bottomrule
    \end{tabular}
 \end{threeparttable}
\end{table*}

\begin{table*}[t]
\renewcommand\arraystretch{1.7}
\centering
\begin{threeparttable}
\caption[]{}\label{tab: reg2} 
  \begin{tabular}{c|c|c|c|c|c|c}
      \toprule
     Stages                      &   \multicolumn{2}{c|}{Dissipation-free}  &     \multicolumn{2}{c|}{Viscous}    &   \multicolumn{2}{c}{Nonlinear }          \\
    \hline
    \multicolumn{7}{c}{Partially ionized gas, $\mathcal{R}=1$} \\
    \hline
    \multirow{2}*{ $\mathcal{E}$}
    &  $\sim e^{2\Gamma_\nu t}$ & $\mathcal{E}(t_{12})$  & $\sim e^{\frac{1}{3}\Gamma_\nu t}$ & $\mathcal{E}_\text{cr}$  & $\sim t$                           & $\mathcal{E}_\text{sat,nl}$     \\
    &  Eq. \eqref{eq: enedif}         & Eq. \eqref{eq: ene12}  &  Eq. \eqref{eq: enedamr}          &  $E_{k,\nu}$  (Eq. \eqref{eq: kinenu})  &  Eq. \eqref{eq: ennoncr}  &  Eq. \eqref{eq: satnlf}          \\
    \hline
    \multirow{2}*{ $k_p$}
  & $\sim e^{\frac{3}{5} \Gamma_\nu t}$ & $k_d(t_{12})$        & $\sim e^{-\frac{1}{6}\Gamma_\nu t}$ &  $k_\text{cr}$    &  $\sim t^{-\frac{3}{2}}$                   &  $L^{-1}$ \\
  & Eq. \eqref{eq: kpdif}                       & Eq. \eqref{eq: kd12}  & Eq. \eqref{eq: dagdar}                       &  $k_\nu$ (Eq. \eqref{eq: knu})       &      Eq. \eqref{eq: evokpcri}   &                 \\
    \hline
    \multirow{2}*{$t$}
    &                                            &  $t_{12}$                       &       $\tau_\text{vis}$                 & $t_\text{cr}$                 &   $\tau_\text{nl}$               &  $t_\text{sat,nl}$       \\
    &                                            &   Eq. \eqref{eq: tim12}   &      Eq. \eqref{eq: tranvis}         &  Eq. \eqref{eq: tnlhesv}  &   Eq. \eqref{eq: nontdrl}       &  Eq. \eqref{eq: rltfsa}    \\
    \bottomrule
    \end{tabular}
 \end{threeparttable}
\end{table*}

\begin{table*}[t]
\renewcommand\arraystretch{1.7}
\centering
\begin{threeparttable}
\caption[]{}\label{tab: reg3} 
  \begin{tabular}{c|c|c|c|c|c|c|c|c}
      \toprule
     Stages                      &   \multicolumn{2}{c|}{Dissipation-free}  &     \multicolumn{2}{c|}{Viscous}    &     \multicolumn{2}{c|}{Transitional}  &   \multicolumn{2}{c}{Nonlinear }          \\
     \hline
    \multicolumn{9}{c}{Conducting fluid, $1<P_m<5^{\frac{4}{5}} \Big(\frac{E_{k,\nu}}{\mathcal{E}_0}\Big)^\frac{1}{2}$} \\
    \hline
    \multirow{2}*{ $\mathcal{E}$}
    &  $\sim e^{2\Gamma_\nu t}$  & $\mathcal{E}(t_{12})$      & $\sim e^{\frac{3}{4}\Gamma_\nu t}$    & $\mathcal{E}_\text{cr}$  & $\mathcal{E}_\text{cr}$     & $\mathcal{E}_\text{cr}$ &  $\sim t$                                &$\mathcal{E}_\text{sat,nl}$    \\
    &  Eq. \eqref{eq: enedif}           &  Eq. \eqref{eq: hpet12}   & Eq. \eqref{eq: hpevis}                         &   $E_{k,\nu}$ (Eq. \eqref{eq: kinenu})  &  $E_{k,\nu}$ (Eq. \eqref{eq: kinenu})  &  $E_{k,\nu}$ (Eq. \eqref{eq: kinenu})   &  Eq. \eqref{eq: ennoncr}     & Eq. \eqref{eq: satnlf}  \\
    \hline
     \multirow{2}*{ $k_p$}
& $\sim e^{\frac{3}{5} \Gamma_\nu t}$ &  $k_R$  & $k_R$  & $k_R$ &  $\sim e^{-\frac{3}{10}\Gamma_\nu t}$ &  $k_\text{cr}$  &  $\sim t^{-\frac{3}{2}}$                   &  $L^{-1}$ \\
& Eq. \eqref{eq: kpdif}                          & Eq. \eqref{eq: hpkr}  &  Eq. \eqref{eq: hpkr} &  Eq. \eqref{eq: hpkr} &  Eq. \eqref{eq: fnonkp}                        & $k_\nu$ (Eq. \eqref{eq: knui})  &  Eq. \eqref{eq: evokpcri}   &                 \\
    \hline
    \multirow{2}*{$t$}
    &                                              &  $t_{12}$                         &  $\tau_\text{vis}$                                & $t_\text{sat,k}$                     & $\tau_\text{tran}$    &  $t_\text{cr}$        &   $\tau_\text{nl}$               &  $t_\text{sat,nl}$       \\
    &                                              &  Eq. \eqref{eq: hpt12}      &  Eq. \eqref{eq: hptauvis}                     &  Eq. \eqref{eq: hpspcrt}  & Eq. \eqref{eq: ttranlp} &  Eq. \eqref{eq: tnlhesv}  &   Eq. \eqref{eq: nontdrl}       &  Eq. \eqref{eq: rltfsa}    \\ 
    \hline
    \multicolumn{9}{c}{Partially ionized gas, $1<\mathcal{R}<5^{\frac{4}{5}} \Big(\frac{E_{k,\nu}}{\mathcal{E}_0}\Big)^\frac{1}{2}$} \\
    \hline
    \multirow{2}*{ $\mathcal{E}$}
    &  $\sim e^{2\Gamma_\nu t}$ & $\mathcal{E}(t_{12})$  & $\sim e^{\frac{1}{3}\Gamma_\nu t}$ & $\mathcal{E}_\text{cr}$  & $\mathcal{E}_\text{cr}$   & $\mathcal{E}_\text{cr}$   & $\sim t$        &  $\mathcal{E}_\text{sat,nl}$    \\
    &  Eq. \eqref{eq: enedif}         & Eq. \eqref{eq: ene12}  &  Eq. \eqref{eq: enedamr}          & $E_{k,\nu}$ (Eq. \eqref{eq: kinenu}) &  $E_{k,\nu}$ (Eq. \eqref{eq: kinenu}) &  $E_{k,\nu}$ (Eq. \eqref{eq: kinenu}) &  Eq. \eqref{eq: ennoncr}  & Eq. \eqref{eq: satnlf}  \\
    \hline
    \multirow{2}*{ $k_p$}
& $\sim e^{\frac{3}{5} \Gamma_\nu t}$ & $k_d(t_{12})$            & $\sim e^{-\frac{1}{6}\Gamma_\nu t}$ & $\mathcal{R}^\frac{1}{2} k_\nu$  & $\sim e^{-\frac{3}{10}\Gamma_\nu t}$ &  $k_\text{cr}$  &  $\sim t^{-\frac{3}{2}}$                   &  $L^{-1}$ \\
& Eq. \eqref{eq: kpdif}                          & Eq. \eqref{eq: kd12}  & Eq. \eqref{eq: dagdar}                       &   Eq. \eqref{eq: hikdve}                &  Eq. \eqref{eq: fnonkp}                        & $k_\nu$ (Eq. \eqref{eq: knu})   &  Eq. \eqref{eq: evokpcri}    &                 \\
    \hline
    \multirow{2}*{$t$}
    &                                            &  $t_{12}$                       &          $\tau_\text{vis}$              & $t_\text{sat,k}$                 &    $\tau_\text{tran}$              &  $t_\text{cr}$        &   $\tau_\text{nl}$               &  $t_\text{sat,nl}$       \\
    &                                            &   Eq. \eqref{eq: tim12}   &          Eq. \eqref{eq: rltvis}        &  Eq. \eqref{eq: tcrrla}   &  Eq. \eqref{eq: ttranlr}      &  Eq. \eqref{eq: tnlhesv}  &   Eq. \eqref{eq: nontdrl}       &  Eq. \eqref{eq: rltfsa}    \\ 
    \bottomrule
    \end{tabular}
 \end{threeparttable}
\end{table*}

\begin{table*}[t]
\renewcommand\arraystretch{1.7}
\centering
\begin{threeparttable}
\caption[]{}\label{tab: reg5} 
  \begin{tabular}{c|c|c|c|c|c|c}
      \toprule
     Stages                      &   \multicolumn{2}{c|}{Dissipation-free}  &  \multicolumn{2}{c|}{Transitional}   &   \multicolumn{2}{c}{Nonlinear }          \\
    \hline
     \multicolumn{7}{c}{ 
     Conducting fluid, $P_m\geq 5^{\frac{4}{5}} \Big(\frac{E_{k,\nu}}{\mathcal{E}_0}\Big)^\frac{1}{2}$ ~~~~
     Partially ionized gas, $\mathcal{R}\geq 5^{\frac{4}{5}} \Big(\frac{E_{k,\nu}}{\mathcal{E}_0}\Big)^\frac{1}{2}$ }  \\
     \hline
    \multirow{2}*{ $\mathcal{E}$}
    &  $\sim e^{2\Gamma_\nu t}$  & $\mathcal{E}_\text{cr}$  & $\mathcal{E}_\text{cr}$ & $\mathcal{E}_\text{cr}$ & $\sim t$                     &  $\mathcal{E}_\text{sat,nl}$    \\
    &  Eq. \eqref{eq: enedif}           &  $E_{k,\nu}$ (Eq. \eqref{eq: kinenu})  & $E_{k,\nu}$ (Eq. \eqref{eq: kinenu}) & $E_{k,\nu}$ (Eq. \eqref{eq: kinenu}) & Eq. \eqref{eq: ennoncr}  &  Eq. \eqref{eq: satnlf}  \\
    \hline
    \multirow{2}*{$k_p$}
  & $\sim e^{\frac{3}{5} \Gamma_\nu t}$  & $k_p (t_\text{sat,k})$     &  $\sim e^{-\frac{3}{10}\Gamma_\nu t}$ &  $k_\text{cr}$  &  $\sim t^{-\frac{3}{2}}$                   &  $L^{-1}$ \\
  & Eq. \eqref{eq: kpdif}                            &  Eq. \eqref{eq: hrdfkps}   &  Eq. \eqref{eq: fnonkp}                        & $k_\nu$ (Eq. \eqref{eq: knui} / \eqref{eq: knu})   &  Eq. \eqref{eq: evokpcri}    &                 \\
    \hline
    \multirow{2}*{$t$}
    &                                              & $t_\text{sat,k}$               & $\tau_\text{tran}$           &  $t_\text{cr}$        &   $\tau_\text{nl}$               &  $t_\text{sat,nl}$       \\
    &                                              &  Eq. \eqref{eq: dfrtcr}      & Eq. \eqref{eq: trantlr}    &  Eq. \eqref{eq: tnlhesv}  &   Eq. \eqref{eq: nontdrl}       &  Eq. \eqref{eq: rltfsa}    \\ 
   \bottomrule
    \end{tabular}
 \end{threeparttable}
\end{table*}

\section{Properties of the MHD turbulence developed during the nonlinear stage of turbulent dynamo}

Hydrodynamic turbulence acts to amplify the magnetic energy. Its
magnetic counterpart is Alfv\'{e}nic turbulence that is described by the 
\citet{GS95}
theory 
(see \cite{Brad13} for a review).
Alfv\'{e}nic turbulence is driven at the equipartition scale during the nonlinear stage, 
where the transition from hydrodynamic turbulence to MHD turbulence occurs. 
Along with the increase of equipartition scale, the domain of Alfv\'{e}nic turbulence expands with time and 
eventually spreads over the entire inertial range of hydrodynamic turbulence. 
The cascade of Alfv\'{e}nic turbulence results from the nonlinear interactions of Alfv\'{e}n perturbations and 
accounts for the universal efficiency of the nonlinear stage.

The incompressible MHD turbulence considered in this paper is actually the Alfv\'{e}nic turbulence. 
In realistic compressible turbulent medium, the MHD turbulence can be presented as a superposition 
of the cascades of Alfv\'{e}nic, slow, and fast modes 
\citep{LG01,CL02_PRL,CL03,KL10}.
The nonlinear interactions with the compressible modes only marginally affect the Alfv\'{e}nic cascade
\citep{CL03} 
and thus our analysis on the turbulent dynamo is also applicable in the 
presence of slow and fast modes in realistic compressible astrophysical fluids.

As the fundamental ingredient of the turbulent dynamo, it is instructive to discuss the properties of MHD turbulence in the dynamo context.

\subsection{Relation between the transitional stage and viscosity-dominated MHD regime}

The transitional stage emerges
for the turbulent dynamo at $P_m>1$ in a conducting fluid and 
$\mathcal{R}>1$ in partially ionized gas.
This criterion for the appearance of the transitional stage in the case of a conducting fluid can be easily tested by numerical simulations. 
We take the numerical results from 
\citet{Bran05}
as an example. 
It is clear that the magnetic spectral tail in the sub-viscous range is absent in the simulation with $P_m=1$,
but present in the simulation with $P_m = 50$ (see Fig. \ref{fig: sket}).
Upon the saturation of the kinematic dynamo in the deep sub-viscous range, 
the Lorentz back-reaction on smaller scales ($k>k_p$) induces fluid motions 
to counteract and suppress the velocity shear. 
The balance between the magnetic tension force and viscous force is established during this process. 
Consequently, the spectral peak shifts to larger scales and the initial Kazantsev spectrum is deformed.
At the end of the transitional stage, the balance between the magnetic energy and kinetic energy settles in the whole 
sub-viscous range from the hydrodynamic viscous scale to magnetic dissipation scale.
The resultant magnetic energy spectrum peaks at the viscous scale and has 
a negative slope as $k^{-1}$ over the sub-viscous scales.
It persists in the following nonlinear stage as long as the dissipation scale is below the viscous scale. 
The simulated turbulent dynamo action in the case of $P_m>1$ by 
\citet{Hau04} 
indicates the $k^{-1}$ subrange for the magnetic spectrum below the viscous cutoff following a $k^{-5/3}$ range of spectra for both kinetic and magnetic energies.
Besides in the context of turbulent dynamo, 
the same power-law tail below the viscous cutoff was also encountered in the viscosity-dominated regime of MHD turbulence
with imposed large-scale magnetic field
\citep{CLV_newregime, CLV03, LVC04}.

The magnetic structure in the viscous-damped region is created by the shear from the viscous-scale eddies and evolves as 
a result of the balance between magnetic tension force and viscous drag.  
The numerical simulations of the kinematic dynamo over the sub-viscous range by 
\citet{Sc02, Schek02}
show
a folding structure of magnetic fields with the length comparable to the viscous scale and thickness of the resistive scale. 
At the end of the transitional stage, the saturated spectral form $k^{-1}$ 
peaks at the viscous scale. This is consistent with the folding structure 
in view of its viscous-scale coherence in the direction parallel to the local magnetic field.

Our analysis for the kinematic stage of turbulent dynamo shows that 
starting from a viscous-scale fluctuation, the bulk of magnetic energy first propagates toward smaller scales until reaching the dissipation scale 
(dissipation-free and viscous stages),
but then moves toward larger scales and back to the viscous scale (transitional stage). 
The emergence of the transitional stage is crucial for properly determining the saturation state of the kinematic stage and 
provides the necessary conditions for the onset of the nonlinear stage.

\subsection{Magnetic reconnection in the kinematic and nonlinear regimes of dynamo}

As mentioned above, 
the numerical simulations of the kinematic dynamo carried out by 
\citet{Schek02, Sc02}
revealed a folding structure of magnetic fields. 
The sheetlike configuration of laminar magnetic fields allows the Sweet-Parker magnetic reconnection 
\citep{Park57,Swe58}
to take place, with the sheets of folded fields separated by current sheets 
\citep{GS06}.
The thickness of the current sheets are determined by magnetic diffusivity. 
Below the scale of the smallest turbulent eddies, it is given by the resistive scale in fully ionized gas and ion-neutral collisional damping
scale in partially ionized gas. 
Within the inertial range of MHD turbulence, the turbulent diffusion of magnetic fields dominates over microscopic magnetic diffusion processes. 
At the equipartition scale, the turbulent diffusion rate is comparable to the rate of stretching by the turbulent eddies, 
and thus further stretching toward thinner current sheets below the equipartition scale is suppressed.

As regards the MHD turbulence developed in the nonlinear stage of turbulent dynamo, 
the condition for the Sweet-Parker reconnection is violated due to the effect of turbulent diffusion. 
Instead, turbulent reconnection of magnetic fields emerges as a natural 
consequence as well as the origin of the turbulent diffusion of magnetic fields
(\cite{LV99,Eyink2011}, see review by \cite{La15}).
The rapid magnetic reconnection between adjacent turbulent eddies within every eddy turnover time
releases the magnetic tension and enables turbulent motions of fluid amidst equipartition magnetic fields, which are otherwise restricted to 
oscillating motions only. 
As a result, both wavelike and turbulent motions exist in the dynamo-generated magnetic fields. 
Their coupling relation is described by a critical balance between the parallel and perpendicular motions of an eddy in MHD turbulence
\citep{GS95},
which is the equality between the period of Alfv\'{e}nic waves over the eddy's parallel scale and 
the eddy turnover time. 
Notice that the parallel and perpendicular scales of eddies should be measured with respect to the local magnetic field
\citep{LV99,MG01,CLV_incomp}.
Moreover, turbulent reconnection in MHD turbulence provides the necessary diffusion rate to prevent the magnetic field from creating 
unresolved knots in the local magnetic field lines. 
Such tangled magnetic field would be inhibitive to turbulent motions and have a shallow magnetic spectrum 
with a significant excess of magnetic energy at small scales
\citep{LV99}.

\subsection{Damping of MHD turbulence in the 
nonlinear stage in partially ionized gas}

With respect to the MHD turbulence in partially ionized gas, 
\citet{LVC04} 
point out that the new regime of MHD turbulence only occurs 
at a relatively high ionization fraction when the ion-neutral collisional damping is subdominant compared with 
the damping due to neutral viscosity.  
This condition agrees with the criterion $\mathcal{R}>1$, i.e., $\xi_i>\xi_{i,\text{cr}}$, which 
guarantees the presence of the transitional stage of the kinematic dynamo in partially ionized gas. 
If we look into the expression of $\mathcal{R}$, 
with the substitution $\Gamma_\nu = k_\nu^2 \nu_n$, Eq. \eqref{eq: rreaex} can be written as 
\begin{equation}
   \mathcal{R} = \frac{k_\nu^2 \nu_n}{k_\nu^2E_{k,\nu}\mathcal{C}},
\end{equation}
which is the ratio between the viscous damping rate and ion-neutral collisional damping rate corresponding to 
$E_{k,\nu}$ at the viscous scale. 

Both neutral viscosity and ion-neutral collisions act as damping effects of the MHD turbulence
generated during the nonlinear stage. Due to the scale-dependent turbulence
anisotropy developed along the MHD cascade, the ratio between the two damping rates in the range of MHD turbulence
has a dependence on $k$,
\begin{equation}\label{eq: radamr}
 r = \frac{k^2 \nu_n}{\omega_d} = \frac{\nu_n}{ 3 \mathcal{C} \mathcal{E}} \frac{k^2}{k_\|^2}.
\end{equation}
According to the 
\citet{GS95}
scaling relation of MHD turbulence, the wavenumbers
parallel and perpendicular to the local magnetic field direction are related by
\begin{equation}\label{eq: ppsrgs}
    k_\| \sim k_p^\frac{1}{3} k_\perp^\frac{2}{3}.
\end{equation}
Here the equipartition scale $k_p$ is considered as the driving scale of MHD turbulence. 
The magnetic energy $\mathcal{E}$ in Eq. \eqref{eq: radamr} is given by Eq. \eqref{eq: enenli}.
Together with Eq. \eqref{eq: ppsrgs}, $r$ is cast into the form 
\begin{equation}\label{eq: ratdyn}
   r =    \frac{2\nu_n}{ 3 \mathcal{C} } L^\frac{2}{3} V_L^{-2} k^\frac{2}{3}, 
\end{equation}
which increases toward smaller scales. 
Here the assumption of strong turbulence anisotropy $k_\perp \sim k $ is adopted for simplicity. 
It applies at sufficiently small scales where the magnetic field plays a dynamically prominent role and 
turbulent eddies are strongly elongated along the local magnetic field direction. 
We see that the dependence on magnetic energy in Eq. \eqref{eq: radamr} vanishes after the scaling relation 
Eq. \eqref{eq: ppsrgs} is taken into account, and hence $r$ is stationary in time.

The scale with comparable damping rates is set by $r=1$, 
\begin{equation}
   k_\text{r=1} = \Big(\frac{ 3 \mathcal{C} } {2\nu_n}\Big)^\frac{3}{2} L^{-1} V_L^3, 
\end{equation}
which can also be equivalently expressed in terms of $\mathcal{R}$, 
\begin{equation}
   k_\text{r=1} = 3^\frac{3}{2} \mathcal{R}^{-\frac{3}{2}} k_\nu.
\end{equation}
At $\mathcal{R}\le1$, $k_\text{r=1}$ lies in the sub-viscous range, 
so that $r<1$ holds and ion-neutral collisional damping dominates neutral viscous damping 
in the entire inertial range of MHD turbulence. 
But when $\mathcal{R}$ is large, $k_\text{r=1}$ can be substantially reduced, 
and the Alfv\'{e}nic cascade can be truncated at the viscous scale.

Both damping processes of MHD turbulence with the turbulent energy injected at a large scale have been
studied quite thoroughly in e.g., 
\citet{LG01, LVC04, XLY14}. 
By analytically solving the dispersion relation of Alfv\'{e}n waves,
\citet{XLY14}
also obtained the ratio between the two damping rates, and provided its 
varying expressions in different regimes of MHD turbulence. 
As to the MHD turbulence arising in the nonlinear stage of turbulent dynamo, 
Eq. \eqref{eq: ratdyn} corresponds to damping of trans-Alfv\'{e}nic turbulence, or the 
\citet{GS95}
type of turbulence.

\section{Application to the formation of the first stars and first galaxies}

\begin{table*}[t]
\renewcommand\arraystretch{2}
\centering
\begin{threeparttable}
\caption[]{The parameters adopted for the first stars and first galaxies
}\label{tab:para} 
  \begin{tabular}{ccccccc}
      \toprule
                                &    $L$ [pc]   & $V_L$  [km s$^{-1}$]    &   T [K]     & $n$ [cm$^{-3}$]  & $\xi_i$  & $B_0$ [G]  \\
    \hline
   First star              &       $360$       &   $3.7$                   & $1000$      &  $1$  & $2\times 10^{-4}$  & $10^{-20}$  \\
   First galaxy         &        $100$       &   $20$                    & $5000$      &  $10$  &   $10^{-4}$   & $10^{-20}$  \\
\bottomrule
    \end{tabular}
 \end{threeparttable}
\end{table*}

The theory presented above is developed for an arbitrary $P_m$ and ionization fraction. 
In application to the formation of the first stars and first galaxies, we restrict ourselves to
the turbulent dynamo in a weakly ionized medium.

\subsection{The first stars}

The first stars formed during the collapse of primordial halos. We adopt the parameters for the initial condition following earlier works, e.g., 
\citet{SchiBan10}, \citet{SchoSch12},
and are listed in Table \ref{tab:para}. 
The primordial gas is neutral dominated with a low ionization degree.
We consider the temperature $T$, total number density $n$, and ionization fraction $\xi_i$ as constants during the whole process to simplify the problem. 
The driving scale of turbulence is taken as the thermal Jeans length, $L = \sqrt{\gamma k_B T / (G m_H^2 n)}$, 
and turbulent velocity at $L$ is the sound speed $V_L = \sqrt{\gamma k_B T /m_H}$, with the adiabatic index $\gamma$,
gravitational constant $G$, hydrogen mass $m_H$, and Boltzmann constant $k_B$.
The initial field strength $B_0$ is chosen to have a conservative value 
\citep{Bierm50, Xu08, Laz92, SchoSch12}.
We adopt the drag coefficient as $\gamma_d=3.5\times10^{13}$cm$^3$g$^{-1}$s$^{-1}$ from 
\citet{Drai83}, and $\sigma_{nn}= 10^{-14}$ cm$^2$ as suggested by e.g., 
\citet{VK13}.

We note that in disregard of the magnetic field amplification by gravitational compression, here we only focus on the growth of the magnetic field by turbulent dynamo.
Based on the analysis established in Section \ref{sec: partiond} for a partially ionized gas, we first determine $\mathcal{R}=0.06$ from 
Eq. \eqref{eq: ratgam}. Accordingly, following the expressions summarized in Table \ref{tab: reg1} and using the relation $B= \sqrt{8\pi\rho \mathcal{E}}$, 
Table \ref{tab: fst} presents 
the time dependence of field strength (column 1), time (column 2), spatial scale where the magnetic 
energy spectrum peaks (column 3), and field strength (column 4) at the end of each evolutionary stage. 
Furthermore, with the expressions from Table \ref{tab: reg1} and parameters from Table \ref{tab:para} used, 
Fig. \ref{fig: fsb} and \ref{fig: fss} illustrate the time evolution of $B$ and the peak scale of magnetic energy spectrum $l_p = 1/k_p$.
As is shown, 
the dynamo action during the primordial star formation proceeds in four stages with various behaviors of magnetic field growth 
and changes of the advancing direction of the spectral peak. 
In the dissipation-free stage, magnetic energy is the most efficiently amplified with the highest growth rate, 
but the timescale involved and peak scale of magnetic energy distribution are marginal. 
Going through the other stages in the kinematic stage, as the dissipation of magnetic energy due to ion-neutral collisional damping becomes more significant, 
the efficiency of magnetic field amplification decreases and, as a result, the timescales of the later stages increase.
During the nonlinear stage, the field strength grows with a square root of time dependence resulting from the back-reaction of strong magnetic field. 
The timescale of the turbulent dynamo is actually determined by that of the nonlinear stage.

The magnetic field is dramatically amplified from the initial seed field of $10^{-20}$ G to the saturated field strength of $\sim 10^{-6}$ G at the end of the dynamo. 
Starting from the viscous stage, magnetic energy initially accumulated at small scales is transferred to ever-larger scales up to the Jeans scale. 
The separate contributions of the kinematic and nonlinear stages in amplifying the magnetic field and transferring 
the magnetic energy toward large scales can also be seen from Table \ref{tab: fst}.
At the end of the kinematic stage, on a timescale of one-tenth of the free-fall time, which is 
\begin{equation}
    t_\text{ff} = \sqrt{\frac{3\pi}{32 G \rho}} = 51.5~ \text{Myr}, 
\end{equation}
the magnetic field strength reaches about $10^{-7}$ G. That is only smaller than the final saturation field strength 
by one order of magnitude. We recall that Eq. \eqref{eq: criene} indicates that a small $\mathcal{R}$ can lead to a high $\mathcal{E}_\text{cr}$. 
It shows that in a weakly ionized medium, namely, $\xi_i < \xi_{i,\text{cr}}$ (Eq. \eqref{eq: crionfrc}), the kinematic stage alone can be sufficient 
for producing the magnetic field strength significantly stronger than the saturated value on the viscous scale 
(see Fig. \ref{fig: fsb})
within a timescale much shorter than the free-fall time. 
However, the kinematic stage is inadequate to account for the large-scale component of the magnetic field. It is the nonlinear stage that brings the bulk 
of magnetic energy over 3 decades in scales up to the Jeans scale. 
The time required for generating the magnetic field coherent on the Jeans scale is longer than $t_\text{ff}$ by one order of magnitude. 
Although additional gravitational compression can further strengthen the magnetic field, it cannot promote the transfer of magnetic energy to the outer scale. 
Hence the magnetic field is unable to moderate the gravitational collapse on large scales.
\footnote{A similar conclusion is true for the magnetic field amplification
within present-day super-Alfv\'{e}nic molecular clouds. 
In such clouds the kinetic energy exceeds the magnetic energy over a broad range of scales. To amplify the magnetic energy up to
equipartition on the scale of cloud size, it requires around 6 turbulent crossing times of the cloud (Eq. \eqref{eq: rst4}),
which is longer than the cloud lifetime of $1-2$ crossing times 
\citep{Elm00}.}
In addition, the ion-neutral collisional damping scale rapidly increases along with the growth of magnetic energy. It moves to 
the critical damping scale at the end of the kinematic stage and increases further in the nonlinear stage. 
Since the magnetic fluctuations are subject to severe damping and truncated on the damping scale, 
the magnetic field is unlikely to influence the fragmentation on small scales either. 
Therefore, according to our results, magnetic fields are not expected to be dynamically important in primordial star formation.

\begin{table*}[t]
\renewcommand\arraystretch{1.5}
\centering
\begin{threeparttable}
\caption[]{The first star,  $\mathcal{R}=0.06$ }\label{tab: fst} 
  \begin{tabular}{c|c|c|c}
      \toprule
       \multirow{2}*{Dissipation-free ($\sim e^{\Gamma_\nu t}$)} & $t_{12}$   & $k_d ^{-1} (t_{12})$ & $B(t_{12})$ \\
 & $5.1\times 10^{-1}$ Myr & $1.3\times 10^{-7}$ pc  & $5.3\times 10^{-13}$ G \\
 \hline
 \multirow{2}*{ Viscous ($\sim e^{\frac{1}{6}\Gamma_\nu t}$)} & $t_{23}$ & $k_\nu^{-1}$ &$B(t_{23})$  \\
 &  $2.1$ Myr & $1.9\times 10^{-3}$ pc  & $7.4\times10^{-9}$ G \\
 \hline
 \multirow{2}*{Damping ($\sim t$)} & $t_\text{cr}$ & $k_{d,\text{cr}}^{-1}$ & $B_\text{cr}$ \\
 & $5.4$ Myr & $1.2\times 10^{-1}$ pc & $1.2 \times 10^{-7}$ G\\
 \hline
 \multirow{2}*{Nonlinear ($\sim \sqrt{t}$)}  & $t_4$ & $L$ & $B_\text{sat,nl}$     \\
 & $6.0 \times 10^2$ Myr  & $3.6\times 10^2$ pc & $1.7\times10^{-6}$ G  \\
    \bottomrule
    \end{tabular}
 \end{threeparttable}
\end{table*}

\begin{table*}[t]
\renewcommand\arraystretch{1.5}
\centering
\begin{threeparttable}
\caption[]{The first galaxy,  $\mathcal{R}=0.006$}\label{tab: fga} 
  \begin{tabular}{c|c|c|c}
      \toprule
       \multirow{2}*{Dissipation-free ($\sim e^{\Gamma_\nu t}$)} & $t_{12}$   & $k_d ^{-1} (t_{12})$ & $B(t_{12})$ \\
 & $1.1\times 10^{-2}$ Myr & $6.0\times10^{-9}$ pc  & $1.2\times10^{-12}$ G \\
 \hline
 \multirow{2}*{ Viscous ($\sim e^{\frac{1}{6}\Gamma_\nu t}$)} & $t_{23}$ & $k_\nu^{-1}$ &$B(t_{23})$  \\
 &  $4.4\times10^{-2}$ Myr & $1.2\times10^{-4}$ pc  & $2.5\times10^{-8}$ G \\
 \hline
 \multirow{2}*{Damping ($\sim t$)} & $t_\text{cr}$ & $k_{d,\text{cr}}^{-1}$ & $B_\text{cr}$ \\
 & $7.3\times10^{-1}$ Myr & $2.5\times10^{-1}$ pc & $3.9\times10^{-6}$ G\\
 \hline
 \multirow{2}*{Nonlinear ($\sim \sqrt{t}$)}  & $t_4$ & $L$ & $B_\text{sat,nl}$     \\
 & $31.1$ Myr  & $1.0\times10^2$ pc & $2.9\times10^{-5}$ G  \\
    \bottomrule
    \end{tabular}
 \end{threeparttable}
\end{table*}

\subsection{The first galaxies}
There exist severe uncertainties concerning the initial conditions for forming the first galaxies
\citep{Bro11}.
The parameters we assume for our model of the first galaxies are also listed in Table \ref{tab:para}. 
They are motived by the numerical simulations by 
\citet{Grei08}, 
where the properties of the first galaxies during the assembly of atomic cooling halos was investigated
(see also \citealt{Schob13}).
We again adopt constant temperature and density as a simplified treatment for an illustrative purpose and 
provide an order of magnitude estimate.

\begin{figure*}[htbp]
\centering
\subfigure[First star]{
   \includegraphics[width=8cm]{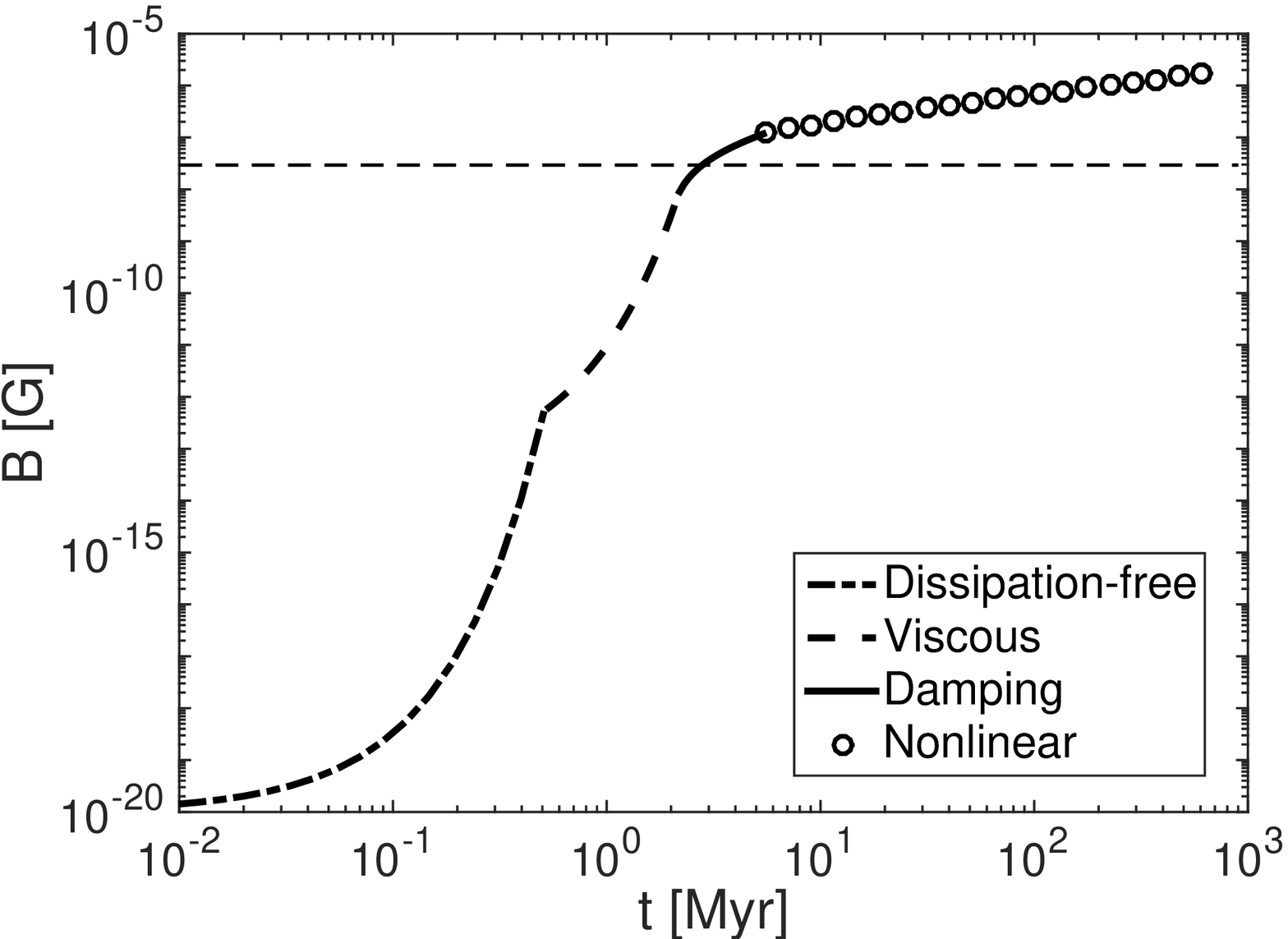}\label{fig: fsb}}
\subfigure[First star]{
   \includegraphics[width=8cm]{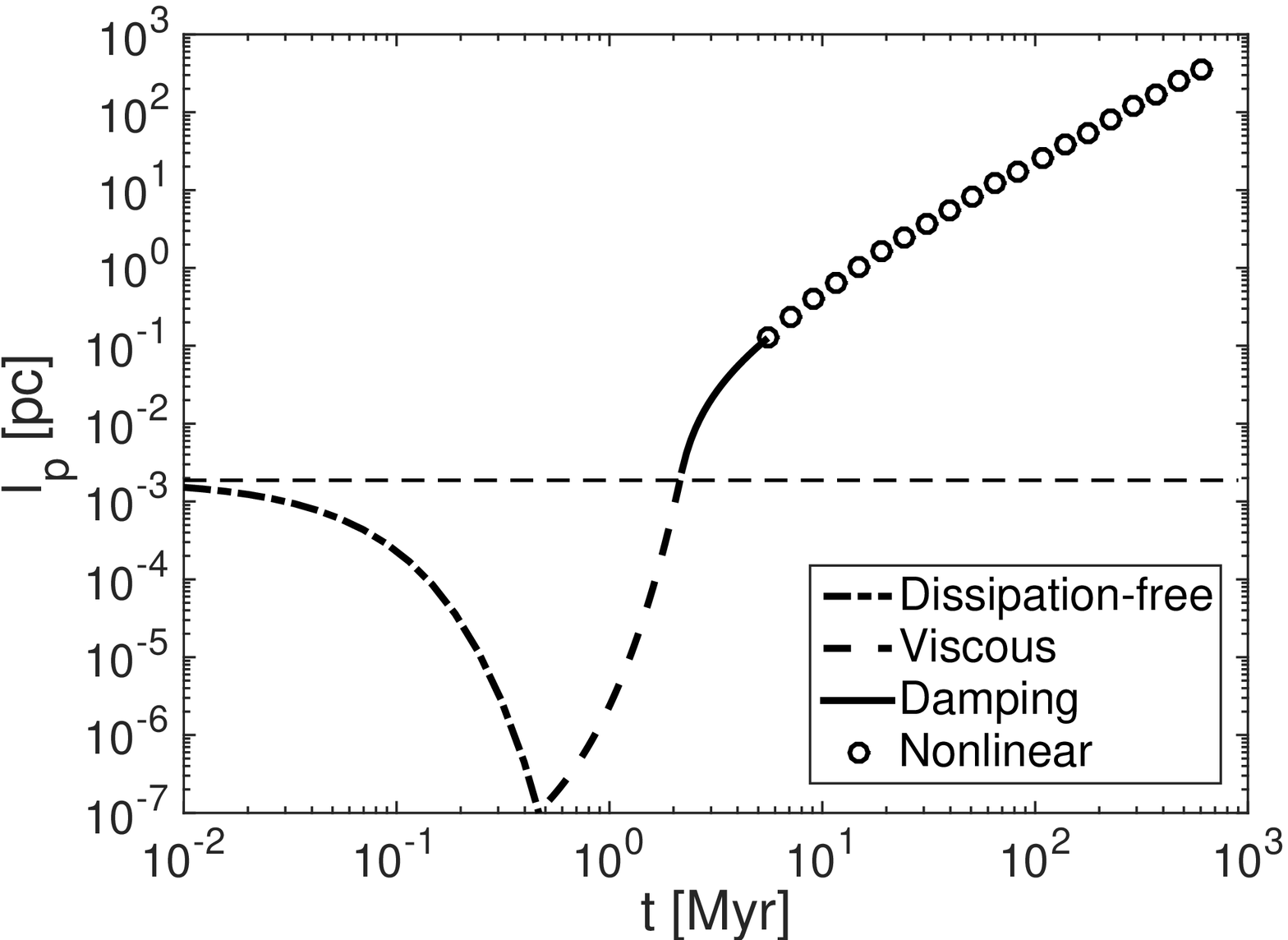}\label{fig: fss}} 
\subfigure[First galaxy]{
   \includegraphics[width=8cm]{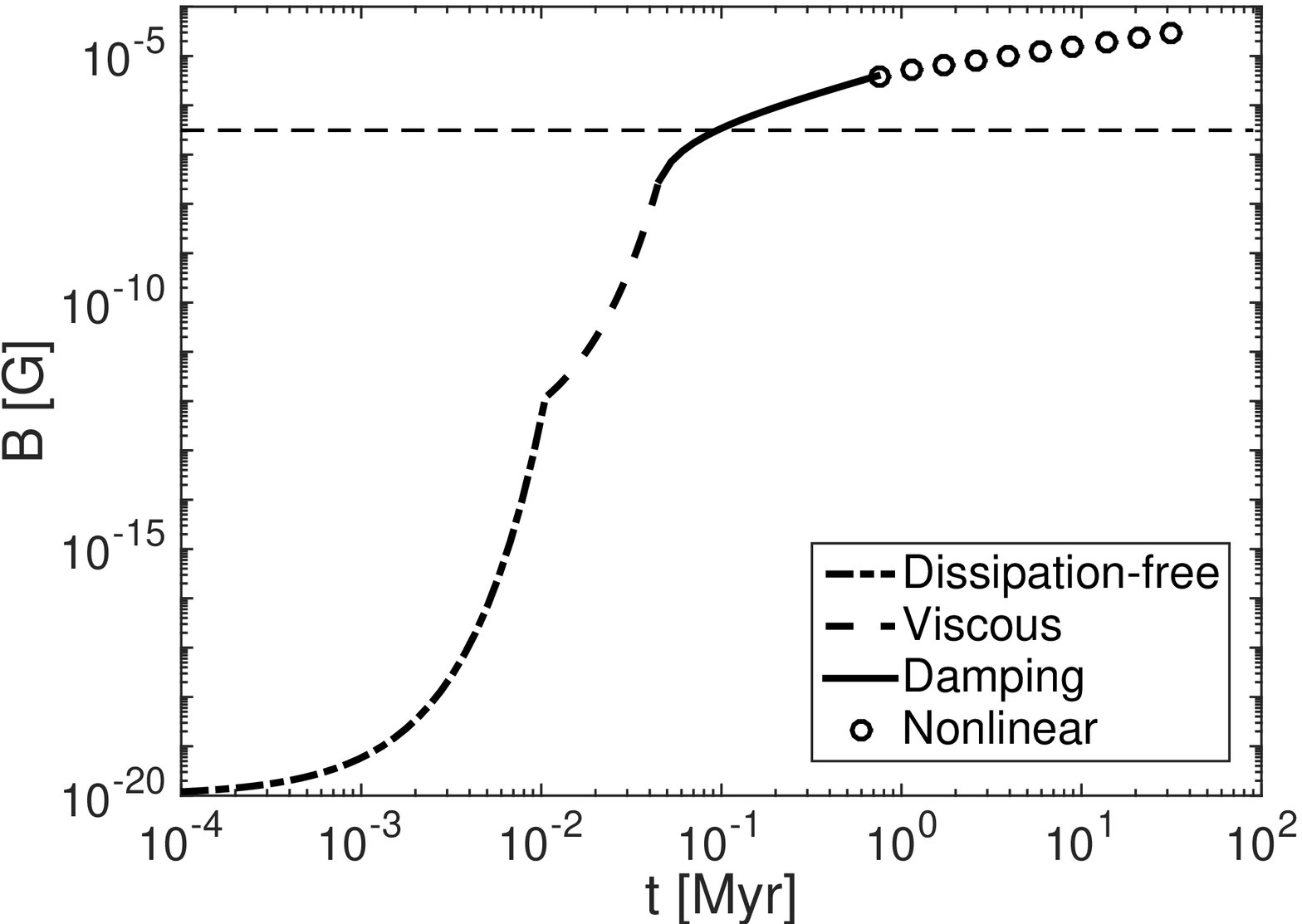}\label{fig: fgb}}
\subfigure[First galaxy]{
   \includegraphics[width=8cm]{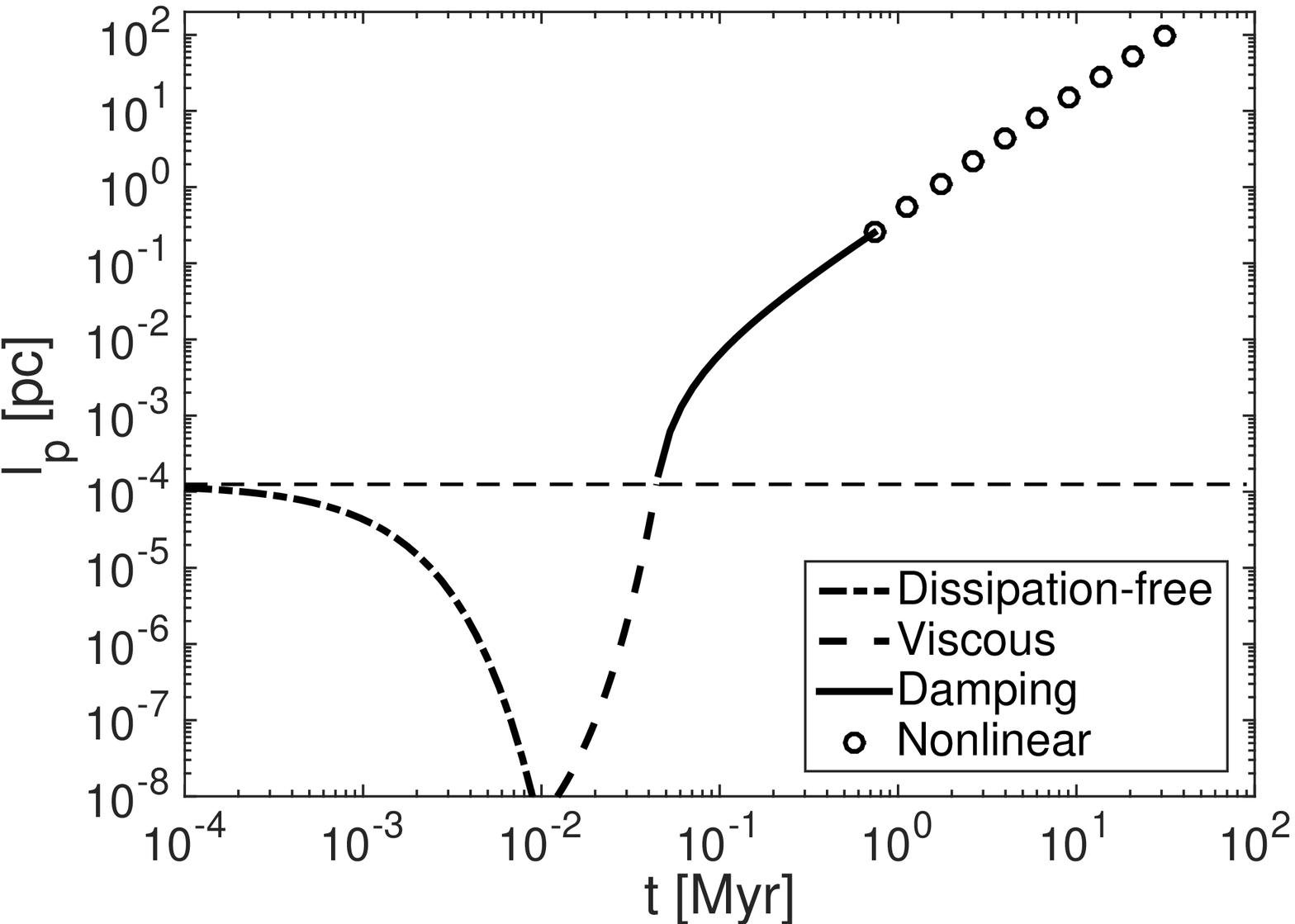}\label{fig: fgs}}  
\caption{ The time evolution of the magnetic field strength and the peak scale of magnetic energy spectrum during the formation of the 
first stars and galaxies. As indicated in the plots, different evolutionary stages are represented by different line styles. 
The horizontal dashed line denotes the magnetic field strength with the corresponding magnetic energy equal to the turbulent energy 
of the viscous-scale eddies in (a) and (c), and the viscous scale in (b) and (d).}
\label{fig: firs}
\end{figure*}

For our model of the first galaxies we find $\mathcal{R}=0.006$. 
The results on the evolution of the magnetic field strength and spectral peak scale 
are displayed in Table \ref{tab: fga}, Fig. \ref{fig: fgb} and \ref{fig: fgs}.
Similar behavior of the turbulent dynamo to that during the formation of the first stars can be observed,
but with a more extended damping stage due to a smaller value of $\mathcal{R}$ (see Fig. \ref{fig: fgb} and \ref{fig: fgs}).
The kinematic stage brings about enormous amplification of the magnetic field, with a strength on the order of $10^{-6}$ G after $7.3\times10^{-1}$ Myr, 
which is negligible compared with the free-fall time $t_\text{ff}=16.3$ Myr.
But the magnetic field coherent at the large turbulence driving scale with the strength $2.9\times10^{-5}$ G can only be reached at a time $31.1$ Myr
after the nonlinear saturation, which is apparently longer than the free-fall timescale. 
It follows that strong magnetic fields with comparable strengths as in local galaxies can be built up in the first galaxies, but the dynamo timescale required 
for the formation of the large-scale galactic magnetic field can be longer than the system's free-fall time, and thus the nonlinear stage can continue 
through the early evolution of the first galaxies.

At the final saturated state the turbulent dynamo is expected to provide magnetic energy comparable to the turbulent energy at the driving scale. 
That is, the level of turbulence determines the asymptotic magnitude of the magnetic field. 
Therefore, the saturated field strengths we obtained are of the order of the resulting large-scale magnetic field in, e.g., 
\citet{SchoSch12, Schob13},
where similar settings for turbulence were used. 
As we mentioned above, 
because of the large uncertainties on the turbulence properties in these primordial environments, it must be kept in mind that the saturation level of 
turbulent dynamo is dependent on the given turbulent condition.

In the gravitationally collapsing primordial gas, additional 
small-scale turbulence can be induced by the collapse in the presence of density inhomogeneities. 
In addition, turbulence can be also amplified by the gravitational collapse, as a result of angular momentum conservation of shrinking eddies. 
Accordingly, the small-scale magnetic field can grow to have a strength stronger than our simple-minded estimate. 
But the fraction of turbulent energy converted to the magnetic energy is still limited by Eq. \eqref{eq: combjl},
which characterize the low efficiency of the turbulent dynamo during the nonlinear stage. 
More detailed study of the turbulent dynamo in the presence of self-gravity will be carried out elsewhere.

Of more importance is the timescale for the saturation on large scales to occur. 
Because of the low efficiency of the nonlinear stage, which was earlier numerically measured
\citep{CVB09, Bere11}
and is now analytically derived in this work, 
the timescale for final saturation we obtained is significantly longer than that was shown in 
\citet{SchoSch12, Schob13}.
As a result, we reach a different conclusion on the importance of magnetic fields for the formation of the first stars and galaxies. 
Moreover, as for a much more extended dynamo process, 
on the one hand, the turbulent energy from the original gravitational collapse may decay and settle down at a lower level during the nonlinear stage. 
On the other hand, in the case of the first stars, it is more likely that the stars formed 
prior to the nonlinear saturation of dynamo since the timescale for the nonlinear stage is considerably longer than the free-fall time.
Hence one may not expect such strong magnetic fields as indicated in Table \ref{tab: fst} and \ref{tab: fga} can be realized 
in a more realistic situation.

\section{Discussion}

By following the Kazantsev theory for studying the turbulent dynamo, 
we trace the time evolution of magnetic energy, with incorporation of  
both microscopic and turbulent diffusion of magnetic fields.
Without 
invoking a particular form of inverse cascade of magnetic energy, 
the Kazantsev dynamo theory with the turbulent diffusion of magnetic fields taken into account naturally resolves the 
nonlinear stage of turbulent dynamo. 
Our analytically derived dynamo efficiency during the nonlinear stage is supported by present numerical evidence.

It is important to point out that we adopt the Kazantsev spectrum in our calculations because it is based on the 
only analytically transparent model of turbulent dynamo. But in fact, 
the exact slope of magnetic energy spectrum does not affect the qualitative behavior of magnetic energy growth in all evolutionary stages. 
That is, the $k^{-1}$ spectrum in the sub-viscous range arising in the transitional stage, 
the linear-in-time growth of $B$ in the damping stage in a weakly ionized gas and 
of $\mathcal{E}$ in the nonlinear stage remain the same when a different slope of the magnetic energy spectrum is given. 
For instance, 
\citet{Eyi10}
(see also \citealt{Krai67})
derived a much steeper slope $4$ instead of $3/2$ as used in this work (Eq. \eqref{eq: mspgrw}) for the magnetic energy spectrum 
in the presence of Kolmogorov turbulence over a broad range of scales. 
In this case, one would still expect the same linear growth of magnetic energy in time during the nonlinear stage, but 
with an even smaller growth rate and more prolonged nonlinear stage as a result of the steeper spectral slope.

The transport of magnetic energy toward larger scales until reaching full equipartition with turbulence throughout the inertial range 
has been envisioned and modeled in earlier analytical studies 
\citep{Bie51, Kuls97, Sub99, Sch02},
and supported by numerical works
\citep{Hau04, Bran05,CVB09, Bere11}.
This behavior was also found in a collapsing system in the simulations carried out by 
\citet{Sur12}, 
and further applied in interpreting the generation of coherent magnetic fields on the 
driving scale of turbulence during the formation of the first stars and galaxies by 
\citet{SchoSch12, Schob13}.
However, the exact magnetic spectrum and magnetic field structure in the final saturated state remains a subject of controversy.
The numerical simulations by e.g., 
\citet{Chou01, Schr02, Ma04}
suggest that at the end of the nonlinear stage, magnetic fields are organized in
folds with the characteristic length at the turbulence driving scale and field reversals at the resistive scale.
The magnetic energy spectrum is dominated by the resistive-scale field.
\footnote{It is worthwhile noticing that even for the folded magnetic fields, 
\citet{Schr02}
claimed that their interaction with the Alfv\'{e}nic turbulence may lead to unwinding of the folds and further energy transport to larger scales, 
until eventual saturation with the Alfv\'{e}nic spectrum of magnetic energy peaking at the outer scale of turbulence.}
Apparently, these results are in contradiction with those mentioned above, 
where no indication of the spectral peak at the resistive scale is found 
\citep{Hau04}.
We caution that to have access to reliable numerical results on the nonlinear stage, 
the necessary requirements include: 
(a) sufficient inertial range not suffering from the dissipation effect,
(b) sufficient computational time until the secularly growing magnetic energy reaches final equipartition, 
which according to our calculations takes at least 6 turnover times of the largest eddy (Eq. \eqref{eq: nontdrl}).
The turbulent dynamo scenario discussed in this work can eventually produce the magnetic fluctuations coherent 
at the outer scale of turbulence,
acting as the externally imposed mean field for trans-Alfv\'{e}nic turbulence, i.e., 
\citet{GS95}
type of MHD turbulence, developed through the entire inertial range.  
This finding has profound implications concerning the build-up of the observed galactic field.

Depending on the relative importance of energy growth to energy dissipation on the viscous scale, namely, 
$P_m$ in a conducting fluid and $\mathcal{R}$ in partially ionized gas, 
the magnetic energy exhibits diverse time-evolution properties in the kinematic stage. 
Numerical investigation on the turbulent dynamo at low $P_m$ or 
in two-fluid (ion-neutral) turbulence is challenging.
The existence of low-$P_m$ turbulent dynamo has been verified numerically, but quantitative results 
are difficult to access due to the resolution constraints
\citep{Hau04, Iska07, Sche07}.
Undoubtedly, 
it is necessary and would be advantageous to carry out sufficiently resolved numerical simulations
over a wide range of $P_m$ and ionization fractions
and detailed comparisons between the results from 
our analysis and numerical simulations in future work.

We stress that 
the remarkable feature of the turbulent dynamo in a weakly ionized medium ($\xi_i<\xi_{i,\text{cr}}$, Eq. \eqref{eq: crionfrc}) is that 
the kinematic stage is largely extended with a considerably higher saturation magnetic energy ($\mathcal{E}_\text{cr}$, Eq. \eqref{eq: criene}) 
on a relatively large scale in the inertial range ($k_{d,\text{cr}}$, Eq. \eqref{eq: cridams}) than the turbulent energy on the viscous scale, and 
its damping stage is characterized by a linear growth of magnetic field strength in time. 
In the application to the first stars and galaxies, it shows that during their assemblage, the kinematic stage is able to produce 
a strong magnetic field on the order of $10^{-7}$-$10^{-6}$ G with an amplification timescale 
smaller than the collapse timescale (free-fall time) by over one order of magnitude, coherent on a scale in the middle of the inertial range of turbulence. 
The subsequent nonlinear stage can further amplify the magnetic field to $10^{-6}$-$10^{-5}$ G and carry most magnetic energy to 
the outer scale of turbulence, which can be comparable to the size of the system, depending on the specific driving mechanism of turbulence
\citep{Schob13}.
The timescale of the nonlinear stage, which is also approximately the total timescale of the turbulent dynamo, is longer than the system's free-fall time. 
So our results suggest that despite the high efficiency of the kinematic stage in amplifying magnetic fields in the first stars and galaxies, the turbulent dynamo 
as a whole is inefficient in generating large-scale magnetic fields within the timescale of gravitational collapse. 
Therefore, magnetic fields may not have played a dynamically important role during the formation of the first stars.
This finding has far-reaching consequences in the primordial initial mass function and subsequent cosmic evolution.

Earlier studies devoted to the turbulent dynamo in primordial star formation and young galaxies include 
\citet{SchoSch12} and \citet{Schob13}. 
They treated the magnetic field amplification as a two-phase dynamo action: 
a kinematic phase with only exponential growth of the magnetic field, and a nonlinear phase with the magnetic energy 
transferred from the viscous scale to the turbulence forcing scale. 
However, our analysis reveals a more complicated physical picture that there exists explicit dependence of the dynamo process on the 
ionization fraction in weakly ionized gas. The essential differences are:

(\romannumeral1) 
We identify three different stages exhibiting both exponential and linear growth of magnetic field strength for the dynamo action in 
the kinematic stage.
It turns out that the kinematic stage has a considerably higher saturation level than that on the viscous scale, and thus has a major contribution in 
the resulting field strength in the environments of the first stars and galaxies. 

(\romannumeral2) The peak of magnetic energy spectrum is fixed at the viscous scale in their calculations. Instead, we follow the evolution of the spectral peak, 
which first propagates deep in the sub-viscous region, and then moves back toward and even beyond the viscous scale during the kinematic stage.

(\romannumeral3) 
For the nonlinear stage, they take the fraction of turbulent energy converted to magnetic energy to be of order unity, 
whereas we derive a universal fraction with a much smaller value ($\approx 0.08$, see Section \ref{ssec: nondyn}), 
which is consistent with numerical results in e.g.,
\citet{CVB09},
\citet{Bere11}.

(\romannumeral4) The nonlinear stage in their consideration depends on the specific dissipation mechanism, but we demonstrate that unlike the kinematic 
stage, the dynamo growth in the nonlinear stage evolves in a universal fashion, irrespective of the microphysical damping processes. 

(\romannumeral5) Contrary to the conclusion reached in 
\citet{SchoSch12}, 
because of the low efficiency of the nonlinear stage and strong ion-neutral collisional damping in partially ionized gas, 
we find that magnetic field is insignificant in the primordial star formation process.\footnote{For the stars formed in the magnetized interstellar medium of the first 
galaxies, we do not rule out the possible magnetic regulation on their formation process. }

\section{Summary}
 
We have investigated the magnetic field amplification by turbulent dynamo in both a conducting fluid with different values of $P_m$ and partially 
ionized gas with different ionization fractions which correspond to the parameter $\mathcal{R}$ (Eq. \eqref{eq: ratgam}).
We find a strong similarity between the dependence of dynamo behavior on $P_m$ and $\mathcal{R}$, 
and identify a number of stages of turbulent dynamo with different dynamo efficiencies. 
We highlight the main results as follows.

1. The dynamo growth of the magnetic energy during the kinematic stage
distinctly varies in different ranges of $P_m$ and $\mathcal{R}$.
Unless $P_m / \mathcal{R}$ is sufficiently high 
($\geq 5^{4/5} \sqrt{E_{k,\nu}/\mathcal{E}_0}$),
the kinematic stage has a sensitive dependence on damping processes.

2. The overall efficiency of magnetic field amplification increases with $P_m / \mathcal{R}$ at $P_m / \mathcal{R}<1$. 
It reaches a constant and becomes independent of $P_m / \mathcal{R}$ at $P_m / \mathcal{R} \geq1$.
Compared to the case with $P_m / \mathcal{R} \geq1$, 
more time is distributed to the kinematic stage but less time to the nonlinear stage in the situation with 
$P_m / \mathcal{R} <1$.

3. The kinematic stage in weakly ionized gas has an extended timescale and goes through a damping stage characterized by 
a linear growth of magnetic field strength in time, 
which is a new predicted regime of dynamo that we propose to test by future numerical simulations.
It has a much higher saturated magnetic energy than the viscous-scale turbulent energy.

4. The transitional stage of the kinematic stage emerges at $P_m / \mathcal{R}>1$, wherein the spectral slope turns to 
$k^{-1}$ in the sub-viscous region.
This $k^{-1}$ tail was earlier reported in numerical simulations with and without an imposed uniform mean magnetic field, 
and here we provided explanation for its physical origin. 
During the transitional stage, the bulk of magnetic energy shifts from a sub-viscous scale back to the viscous scale.

5. The nonlinear stage is unaffected by the microscopic diffusion of the magnetic field and magnetic energy dissipation rate. 
By applying the Kazantsev theory to scales larger than the equipartition scale within the inertial range of turbulence, 
we derived both the linear dependence of magnetic energy on time and the 
universal growth rate of magnetic energy as $3/38 \approx 0.08$ of the turbulent energy transfer rate, 
in good agreement with earlier numerical results.

6. In the context of the first stars and galaxies, the kinematic stage is highly efficient and has a major contribution in boosting the field strength 
and acts in concert with the nonlinear stage in carrying magnetic energy toward large scales. 
But the entire timescale of the dynamo amplification is longer than the free-fall timescale.

7. Due to the inefficiency of the nonlinear stage and strong ion-neutral collisional damping, the turbulent dynamo is inadequate in generating 
dynamically important magnetic field during the primordial star formation. 
\\
\\

We thank the referee for his/her constructive comments.
S.X. acknowledges the support from China Scholarship Council during her stay at the University of Wisconsin-Madison.
A.L. acknowledges the support of NSF grant AST-1212096 and NASA grant NNX14AJ53G.
A.L. thanks the Department of Physics of the University of Rio Grande de Norde (Natal) for their hospitality and Humboldt Foundation for the support of his stay in Bochum.

\bibliographystyle{apj.bst}
\bibliography{yan}

\end{document}